\documentclass[pre,twocolumn,nofootinbib]{revtex4}
\usepackage{graphicx}
\usepackage{color}
\usepackage{epsfig}
\usepackage{multirow}
\begin{document}
\title{Understanding the Dynamics of Glass-forming Liquids with Random Pinning 
within the Random First Order Transition Theory}

\author{Saurish Chakrabarty$^{1,2}$}
\email{saurish@icts.res.in}
\author{Rajsekhar Das$^{3}$}
\email{rajsekhard@tifrh.res.in}
\author{Smarajit Karmakar$^{3}$}
\email{smarajit@tifrh.res.in}
\author{Chandan Dasgupta$^{2,4}$}
\email{cdgupta@physics.iisc.ernet.in}
\affiliation{$^1$ International Centre for Theoretical Sciences,
Tata Institute of Fundamental Research,
Shivakote, Hesaraghatta, Hubli, Bangalore, 560089, India,
$^2$ Centre for Condensed Matter Theory, Department of Physics,
Indian Institute of Science, Bangalore, 560012, India,
$^3$ Centre for 
Interdisciplinary Sciences, Tata Institute of 
Fundamental Research, 21 Brundavan Colony, Narisingi, Hyderabad, India, 
 $^4$ Jawaharlal Nehru Centre for Advanced 
Scientific Research, Bangalore 560064, India.}

\begin{abstract}
Extensive computer simulations are performed for a few model glass-forming liquids in both two and three dimensions to study their dynamics when a randomly chosen fraction of particles are frozen  in their equilibrium positions. For all the studied systems, we find that the temperature-dependence of the $\alpha$ relaxation time extracted from an overlap function 
related to the self part of the density autocorrelation function can be explained within the framework of the Random First Order Transition (RFOT) theory of the glass transition.  
We propose a scaling description to rationalize the simulation results and show that our data for the $\alpha$ relaxation time for all temperatures and pin concentrations are consistent with this description. 
We find that the fragility parameter obtained from fits of the temperature dependence of the  $\alpha$ relaxation time to the Vogel-Fulcher-Tammann (VFT) form decreases by almost an order of magnitude as the pin concentration is increased from zero. Our scaling description relates the fragility parameter to the static length scale of RFOT and thus provides a physical understanding of fragility within the framework of the RFOT theory.  
Implications of these findings for the values of the exponents appearing in the RFOT theory are discussed. 
\end{abstract}
\maketitle

\section{Introduction}
The rapid rise of the viscosity of glass forming liquids with decreasing 
temperature remains one of the important unsolved problems 
in condensed matter physics
\cite{09Cav, 11BB, 14KDS, 00Ediger, 05Berthier, 06BBMR, 08BBCGV, 09KDS}.
In the Random First Order Transition (RFOT) theory~\cite{RFOT,RFOT1,RFOT2}, the growth of the viscosity
is attributed to an {\it ideal glass transition} at which the configurational entropy
associated with the multiplicity of amorphous minima of the free energy
goes to zero.  This transition is supposed to take place at the Kauzmann temperature $T_K$ at which the excess entropy of a supercooled liquid
extrapolates to zero.  This temperature is found to be 
close to $T_{VFT}$, the temperature at
which a Vogel-Fulcher-Tammann (VFT) fit to temperature dependence of the viscosity
of a fragile glass-forming liquid predicts a divergence of the viscosity.   Equilibrium and dynamical
properties of liquids near $T_{VFT} \simeq T_K$ can not be studied in experiments or computer simulations
because liquids fall out of equilibrium at substantially higher temperatures. For this reason, the occurrence of the ideal glass transition 
of RFOT theory remains controversial.

The kinetic fragility\cite{angellFragility}, which measures the rate at which the viscosity increases with 
decreasing temperature, is another interesting but poorly understood aspect of the dynamics near the glass transition. 
The microscopic origin of fragility has been investigated in many numerical studies~\cite{SnehaBimanPaper,coslovichFragility1,berthierWittenFragility, 
shiladityaSrikanthFragility,anshulSrikanthFragilityBarrier}. In these studies, attempts 
were made to tune the fragility by changing different parameters of the
system, {\it e.g.} the degree of polydispersity in particle size 
\cite{SnehaBimanPaper}, softness of the
pair potential \cite{shiladityaSrikanthFragility}, the density 
of a system of particles interacting via a Hertzian potential \cite{berthierWittenFragility}, the composition of 
binary glass-forming liquids \cite{coslovichFragility1}, and the height of a narrow repulsive spike in the pair potential   
\cite{anshulSrikanthFragilityBarrier}. In some of these cases 
\cite{SnehaBimanPaper,shiladityaSrikanthFragility}, the fragility was found to 
depend weakly on the tuning parameters, so that no clear conclusion about the origin of  
fragility could be reached. On the other hand, in Refs. 
\cite{coslovichFragility1,berthierWittenFragility, anshulSrikanthFragilityBarrier}, the fragility 
was found to change by substantial amounts and correlations of the fragility
with other thermodynamic quantities were obtained. However, a clear understanding of the microscopic
origin of fragility remains elusive.   

Recently, it has been proposed~\cite{cammarotaPinning} that the ideal glass transition of RFOT may be accessible in experiments and
simulations of systems in which a randomly chosen fraction of the constituent particles are pinned at their
positions in an equilibrium configuration of the liquid. It is argued~\cite{cammarotaPinning} that the temperature $T_K$ at which
the configurational entropy goes to zero in such a system increases as the concentration of pinned
particles is increased, thereby making it possible to equilibrate the liquid near the higher transition temperature.
Results of numerical studies~\cite{13KB,14KC,okim2015} of equilibrium properties of such randomly pinned systems are consistent with the
predictions of Ref.~\cite{cammarotaPinning}. On the other hand, a numerical study~\cite{CKD15} of the dynamics of two model liquids with
random pinning indicates that the VFT temperature $T_{VFT}$ obtained from the temperature-dependence of the $\alpha$ 
relaxation time extracted from the self part of the intermediate scattering function or a related overlap function remains
constant or decreases slightly with increasing pin concentration. It is also shown in Ref.~\cite{CKD15,15CKDpnas} that the $\alpha$
relaxation time defined there remains finite at temperatures close to the values at which the equilibrium calculation
of Ref.~\cite{okim2015} predicts a vanishing of the configurational entropy. These results raise important questions about the
validity of the relation between the configurational entropy and the $\alpha$ 
relaxation time predicted in RFOT theory.  Possible explanations of this discrepancy have been suggested~\cite{15CKDpnas,okim2015PNASReply},
but a clear resolution is not yet available.

In Ref.~\cite{CKD15}, it was also shown 
that the fragility can be tuned very effectively by changing the fraction of 
pinned particles. Pinning decreases the kinetic fragility and by increasing 
the concentration of pinned particles, one can change the fragility by a 
large factor ($\simeq 10$), comparable to experimentally
observed variations of the fragility in different glass forming liquids. Thus, 
random pinning provides a nice  way of tuning the fragility without changing 
the microscopic details of the system. Recently,  random 
pinning of the kind considered here has been realized 
experimentally~\cite{shreyas2014pinning} in two-dimensional colloidal systems. 
This suggests that it will not be difficult to tune the fragility in 
experimental colloidal glasses in the near future. It is, therefore, important 
to understand the reason for the variation of the fragility 
seen~\cite{CKD15} in the dynamics of randomly pinned liquids.

To address some of the outstanding questions mentioned above, we have carried 
out detailed molecular dynamics simulations of four models of glass-forming 
liquids with random pinning. These models include the three-dimensional models 
considered in our earlier study~\cite{CKD15} and two two-dimensional 
models. The dependence of the $\alpha$-relaxation time, obtained from an 
overlap function analogous to the self part of the density autocorrelation
function, on the temperature and the pin concentration in the two-dimensional 
models is found to be qualitatively similar to that in three dimensions. 
It was suggested in Ref.~\cite{okim2015PNASReply} that the relaxation time 
extracted from the {\it full} density autocorrelation function should be 
considered in a check of whether the relaxation time diverges at the points 
at which the configurational entropy vanishes according to the numerical 
results of Ref.~\cite{okim2015}. We have calculated the autocorrelation
function of the full density in one of the three-dimensional models (the 
Kob-Andersen model~\cite{95KA} which was also considered in 
Ref.~\cite{okim2015}) and found that the time scale associated with the decay 
of this function is actually {\it smaller} than that of the time scale 
extracted from the overlap function. Thus, considering the autocorrelation 
function of the full density does not resolve the discrepancy between 
the behavior of the relaxation time and the results for the configurational 
entropy reported in Ref.~\cite{okim2015}.

We show that an alternative description proposed in Ref.~\cite{CKD15}, 
based on the assumption that the configurational entropy of the pinned system 
differs from that of the unpinned system by a multiplicative factor that 
decreases from unity as the concentration of pinned particles is increased 
from zero, allows us to reconcile the results for the dynamics, as well as new 
results for the configurational entropy of the unpinned systems, with the RFOT 
theory. This assumption leads to a scaling description~\cite{CKD15} of the 
dynamics that involves a new static length scale $\xi_p$ associated with the 
effect of random pinning on the $\alpha$ relaxation time. We show that our 
data for the relaxation time for all temperatures and pin concentrations 
considered in our study are consistent with this scaling description.
The relation between the pinning length scale $\xi_p$ and the static length 
scale $\xi_s$ of RFOT theory is studied in detail. This allows us to estimate 
the values of the exponent $\theta$, associated with the dependence of the 
interface free energy on the system size, and the exponent $\psi$ that 
describes the dependence of the time scale on $\xi_s$ in the RFOT theory.
We find that $\theta \simeq d-1$ ($d$ is the spatial dimension), $\psi \simeq 
1$ for the three-dimensional models considered here.  In one of the 
two-dimensional models studied here, we find that $\theta \simeq d-1$, $\psi 
\simeq 0.7$, whereas the value of $\theta$ estimated for the other
two-dimensional model does not satisfy the bound $\theta \leq d-1$.
We also find that the static length scale of the RFOT theory is intimately 
connected to the fragility, thereby providing a rationale for the variation 
of fragility in different glass forming liquids.              

The rest of the paper is organized as follows. First, we define the models 
studied here and describe the methods used in our simulations and calculations 
of quantities of interest. After a brief review of the predictions of the RFOT 
theory for the relationships among different dynamic and thermodynamic 
quantities, such as the relaxation time ($\tau_{\alpha}$), the static length 
scale ($\xi_s$) and the configurational entropy density ($s_c$), we present 
our numerical results and their analysis within the framework of the
RFOT theory using the scaling description proposed in Ref.~\cite{CKD15}. 
The main results of the scaling analysis for the exponents appearing in the 
RFOT theory and the microscopic origin of the fragility parameter are 
discussed. We close with a summary of the main conclusions and their 
implications for the validity of the RFOT theory.
           
\section{Models and Methods}
\noindent{\bf Models: }We have studied four different model glass
forming liquids in this study -- two models each in two and three
dimensions. The first model glass former we study is the well-known Kob-
Andersen  80:20 binary Lennard-Jones mixture in three dimensions. 
Here it is referred to as the {\bf 3dKA model}. The interaction 
potential in this model is given by 
\begin{equation}
V_{\alpha\beta}(r)=4\epsilon_{\alpha\beta}\left[\left(\frac{\sigma_{\alpha\beta}}{r}\right)^{12}-
\left(\frac{\sigma_{\alpha\beta}}{r}\right)^{6}\right],
\end{equation}
where $\alpha,\beta \in \{A,B\}$ and $\epsilon_{AA}=1.0$, $\epsilon_{AB}=1.5$,
$\epsilon_{BB}=0.5$, $\sigma_{AA}=1.0$, $\sigma_{AB}=0.80$,
$\sigma_{BB}=0.88$. The interaction potential is cut off at 
$2.50\sigma_{\alpha\beta}$ and we use a quadratic polynomial to make the 
potential and its first two derivatives smooth at the cutoff distance. 
The temperature range studied for this model is $T \in [0.45, 3.00]$ 
at number density $\rho = 1.20$.  The same model with 65:35 composition
is used in the two dimensional simulations. We refer to  this model as the 
{\bf 2dMKA model}. The 65:35 composition is chosen to make sure that the 
system does not show any crystalline domains and we checked that within our 
maximum simulation time, the system does not form crystallites in any of the 
$32$ statistically independent simulation runs. 
The second model studied is a 50:50 binary mixture with a pairwise 
interactions that falls off with distance as an 
inverse power law,
\begin{equation}
V_{\alpha\beta}(r) = \epsilon_{\alpha\beta}\left(\frac{\sigma_{\alpha\beta}}{r}\right)^n,
\end{equation}
with exponent $n = 10$ (the {\bf 3dR10 model}). The potential is cut off at 
$1.38\sigma_{\alpha\beta}$. We  again use a quadratic polynomial to make the 
potential and its first two derivative smooth at the cutoff. The parameters of 
the potential are: $\epsilon_{\alpha\beta} = 1.0$, $\sigma_{AA} = 1.0$,
$\sigma_{AB} = 1.22$ and $\sigma_{BB} = 1.40$. The temperature range covered 
for this model is $T \in [0.52,3.00]$ at number density $\rho = 0.81$. The
same model with number density $\rho = 0.85$ is studied in two dimensions. 
We will refer to this model as the {\bf 2dR10 model}.
Length, energy and time scales are measured in units of 
$\sigma_{AA}$, $\epsilon_{AA}$ and $\sqrt{\sigma_{AA}^2/\epsilon_{AA}}$. \\

\noindent{\bf Simulation Details: }NVT molecular dynamics simulations are 
performed in a cubic simulation box with periodic boundary conditions in both 
two and three dimensions for all the model systems. We use the modified 
leap-frog algorithm with the Berendsen thermostat to keep the temperature 
constant in the simulation runs. The use of any other thermostat does not 
change the results significantly as we are mostly interested in 
configurational changes in the system instead of momentum correlations. 
The integration time steps used is $dt = 0.005$ in this temperature range. 
Equilibration runs are performed for $\sim 10^8 - 10^9$ MD steps depending 
on the temperature and production runs are long enough to ensure that the 
two-point density correlation function $Q(t)$ (defined below) goes to zero 
within the simulation time. For all the model systems, we have performed 
simulations for pin concentration (defined as the fraction of pinned
particles) in the range $\rho_{pin} \in [0.005, 
0.200]$  for each temperature. For very low temperatures, we were not able to 
equilibrate the system for high pin concentrations because of a dramatic 
increase in the relaxation time.\\

\noindent{\bf Dynamic Correlation Function: } Dynamics is characterized by the 
two point density-density overlap correlation functions $Q(t)$, defined as:
\begin{equation}
Q(t) = \frac{1}{N-N_p}\left[\left<\sum_{i,j=1}^{N-N_p}w\left( |\vec{r}_i(t) - \vec{r}_j(0)|\right)\right>\right],
\end{equation}
where the window function $w(x) = 1.0$ if $x<0.30$, else $w(x) = 0$. This 
choice of window function is made to remove the short-time vibrational 
component from the correlation function. $N_p$ is the number of pinned 
particles in the system and $\langle\ldots\rangle$ implies thermal averaging 
and $[\ldots]$ means averaging over different realizations of the randomly 
pinned particles. We have calculated also the self part of the above 
correlation function, defined as
\begin{equation}
Q_s(t) = \frac{1}{N-N_p}\left[\left<\sum_{i=1}^{N-N_p}w\left( |\vec{r}_i(t) - \vec{r}_i(0)|\right)\right>\right].
\end{equation}
The relaxation time $\tau_\alpha$ was obtained from $Q_s(t)$ using 
$Q_s(\tau_\alpha)=\exp(-1)$.\\
\begin{figure*}
\begin{center}
\includegraphics[width=2.1\columnwidth]{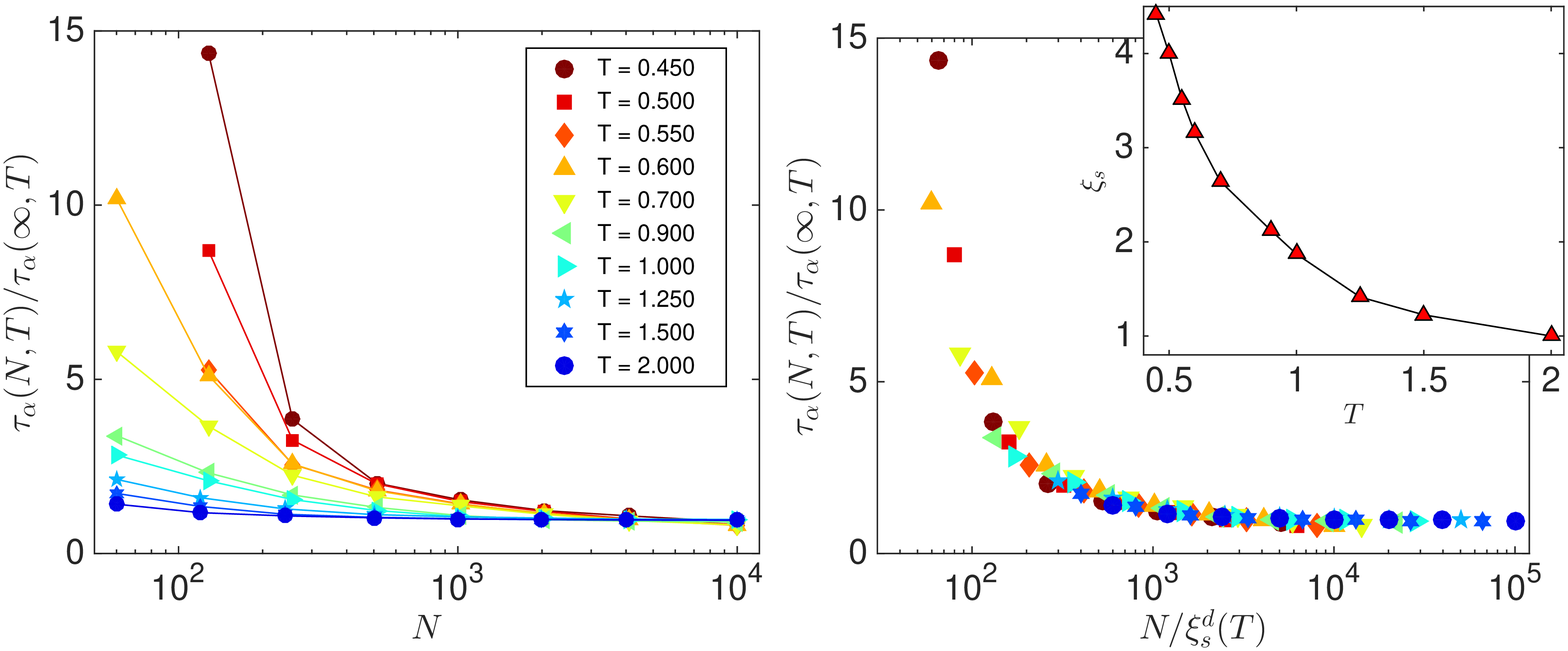}\\
\includegraphics[width=2.15\columnwidth]{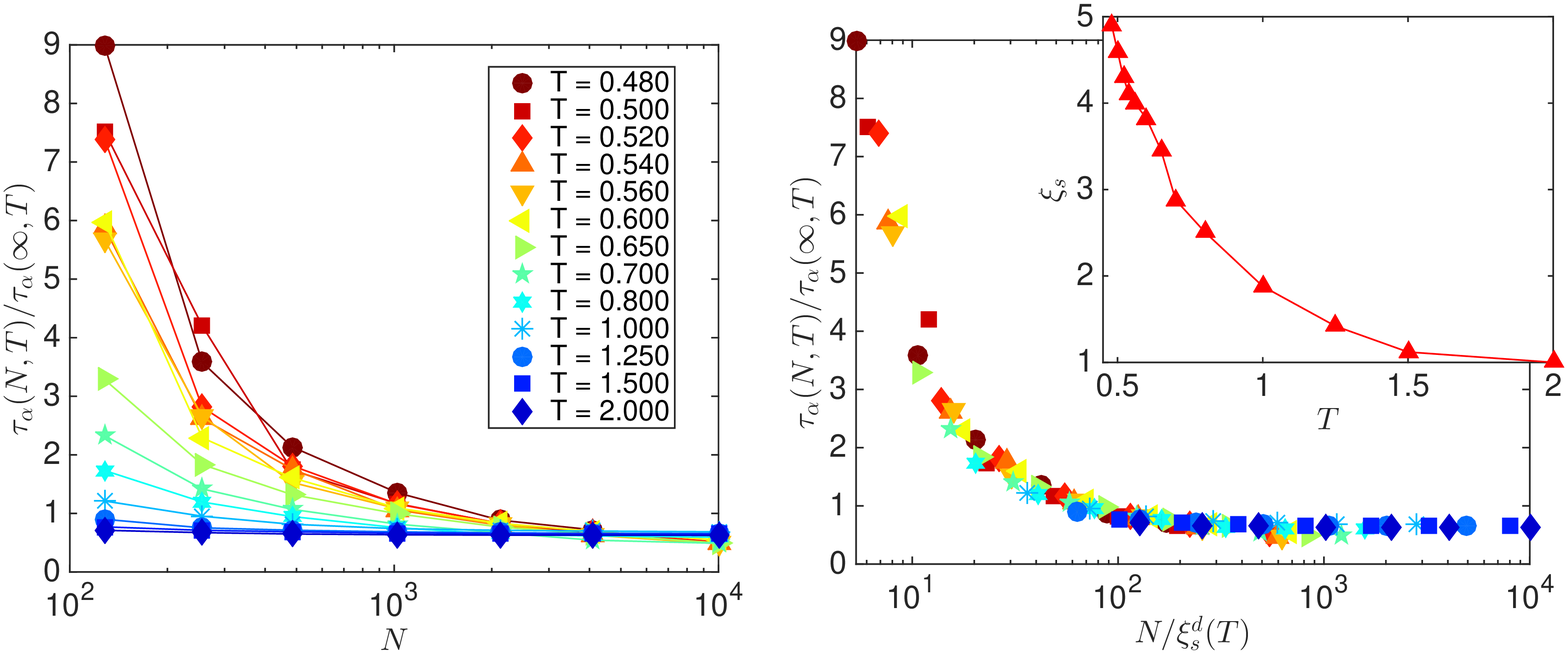}
\caption{Top Panels: Finite-size effects in the $\alpha$ relaxation time of 
the 2dMKA model and the data collapse to obtain the static length scale 
$\xi_s$ (shown in the inset). Bottom Panels: Similar analysis done for the 
2dR10 model with the length scale shown in the inset (see text for details).} 
\label{tauScaling}
\end{center}
\end{figure*}

\noindent{\bf Pinning Protocol: }In this study, we have considered
randomly pinned system in which the pinned particles are chosen randomly 
from an equilibrium configuration of the system at the temperature of 
interest. In \cite{okim2015} a different pinning protocol was used with the 
constraint that the pinning sites should be uniformly distributed in space. 
This, according to \cite{okim2015}, reduces sample to sample fluctuations. 
However, in our study where we used random pinning without any bias, the 
sample to sample fluctuations are found to be similar in magnitude to those 
when the pinning protocol of Ref.~\cite{okim2015} is used. So, we report here 
results obtained from simulations of randomly pinned systems in which the 
pinned particles were chosen without any bias. It is important to mention 
that different choices of the pinning  protocol actually lead to somewhat 
different results, but whether the choice qualitatively changes the physics 
is yet to be understood.

\vskip +0.2cm
\noindent{\bf Configurational Entropy: }
The configurational entropy is calculated here only for the unpinned systems 
and the per particle entropy is defined as 
\begin{equation}
s_c = s_{bulk} - s_{vib}
\end{equation}
where $s_{bulk}$ is the bulk entropy calculated via thermodynamic integration
and $s_{vib}$ is the basin entropy for inherent structures, calculated
using the harmonic approximation\cite{srikanthPRL2000liquidLimits}. Similar 
analysis was recently extended to randomly pinned systems \cite{okim2015}, 
although it is not clear whether the harmonic approximation is sufficiently 
accurate to get reliable estimates of the configurational entropy at the 
higher temperatures considered for the pinned systems \cite{CKD15}.\\ 

\noindent{\bf Static Length Scale: }
The static length scale $\xi_s$ is calculated from a combination of 
finite-size scaling of the minimum eigenvalue and point-to-set (PTS) methods 
as described in detail in Ref.~\cite{BKP13PRL} for the three dimensional 
systems (3dKA and 3dR10). In Ref.~\cite{KDSRoPP}, it was shown that the length scale obtained 
via the methods mentioned above for three dimensional systems completely 
explains the finite size effects seen in the relaxation time $\tau_{\alpha}$ 
for all temperatures including high temperatures. So, for the two dimensional 
systems, we employed the method of finite-size scaling of $\tau_{\alpha}$ 
\cite{12KP} to obtain the static length scales as the eigenvalue method and 
PTS methods involve much larger computational efforts to obtain the length 
scale. In Fig.\ref{tauScaling}, we have shown the finite-size scaling of 
$\tau_{\alpha}$ to obtain the static length scale for both the two 
dimensional models. The data collapses observed are quite good and that give 
us confidence about the reliability of the values of the extracted static 
length scale, $\xi_s$. There is a recent controversy about the 
usefulness of the PTS length scale for two-dimensional glass forming system 
with short-range bond-orientational order \cite{tanakaPNAS2015hexatic}. It is 
claimed that the PTS correlation length is insensitive to the growth of 
structural order in system with pronounced short-range bond-orientational 
order and hence the length scale associated with structural order, which 
governs the relaxation time, is not the same as the PTS length scale. It is 
to be noted that this controversy is not relevant for the two-dimensional 
systems considered here because local bond-orientational order is not 
pronounced in these systems. 
            
\section{Random First Order Transition Theory}
The Random First Order Transition (RFOT) \cite{RFOT, RFOT1, RFOT2}theory is 
motivated by analytic results obtained for infinite-range spin glass models 
whose behavior is qualitatively similar to that of structural glasses. This
theory connects thermodynamics to dynamics via the configurational entropy
$s_c$. The RFOT theory predicts certain relationships connecting the 
$\alpha$-relaxation time $\tau_{\alpha}(T)$, the static length scale 
$\xi_s(T)$ and the configurational entropy $s_c(T)$, with associated
exponents which have not yet been computed from any microscopic
theory. The main ingredient of RFOT theory is the introduction of a surface 
free energy cost for creating an interface between two amorphous states of the 
same glassy system. Unlike conventional liquid-solid interfaces, the free 
energy cost of creating an interface between two amorphous states is somewhat 
difficult to define as there is no obvious order parameter which can 
distinguish between the two amorphous states. Direct measurements of this 
interfacial free energy cost have been attempted in the past, but a clear 
consensus on its dependence on the size of the interface is yet to emerge. 
Thus, a direct measurement of the interfacial free energy and the exponent
that describes the power-law dependence of the free energy on the interface 
size would be very important for verifying the validity of the RFOT theory. In 
this work, we have tried to extract some of this information by introducing
random pinning in the system and measuring the corresponding effects
on the dynamics.  
     
Close to the Kauzmann temperature $T_K$, RFOT theory predicts that 
the static correlation length $\xi_s(T)$ diverges as
\begin{eqnarray}
\xi_s(T) \propto \left[\frac{1}{s_c(T)}\right]^\frac{1}{d-\theta}.
\label{xis}
\end{eqnarray}
where the exponent $\theta$ describes the dependence of the surface free 
energy cost on the linear size of the interface between two amorphous states. 
The $\alpha$-relaxation time is related to the static length-scale as
\begin{equation}
\tau_\alpha = \tau_0\exp{\left[\Delta\xi_s^\psi/T\right]},
\label{tauXi}
\end{equation}
where it is assumed that the typical free energy barrier that has to be  
overcome in order to rearrange a correlated volume of linear size $\xi_s$ 
is $E_a = \Delta\xi_s^{\psi}$. Substitution of Eq.\ref{xis} in 
Eq.\ref{tauXi} leads to a generalized Adam-Gibbs (AG) relation,  
\begin{equation}
\tau_{\alpha}(T) = \tau_0\exp\left[\frac{A}{Ts_c^{\frac{\psi}
{d-\theta}}}\right] 
= \tau_0\exp\left[\frac{A}{Ts_c^{\alpha}}\right],
\label{modAG}
\end{equation}
where $\alpha = \psi/(d - \theta)$. The generalized AG relation, 
Eq.~\ref{modAG}, reduces to the original AG relation if $\psi=d-\theta$. In 
this case, the Vogel-Fulcher-Tammann (VFT) form for the temperature 
dependence of $\tau_\alpha$ is recovered if $Ts_c(T)$ goes to zero at the
Kauzmann temperature $T_K$ as $Ts_c(T) \propto (T - T_K)$. Here we will work 
with the generalized AG relation, Eq. \ref{modAG}, as in \cite{ag-prl}, it was 
shown that the AG relation is violated in two dimensions and the nature of the
deviation from the AG relation depends on the specific details
of the model system studied. For a system that satisfies the generalized AG 
relation with $\alpha \ne 1$, the temperature dependence of $\tau_\alpha$ 
would be described by the VFT equation if $Ts_c^\alpha \propto (T-T_K)$ 
for $T$ close to $T_K$. 

\section{Random Pinning: Scaling arguments} 
In Ref.~\cite{cammarotaPinning}, it was argued that the 
Kauzmann temperature $T_K$ where the configurational entropy $s_c$ goes to
zero should increase with increasing concentration $\rho_{pin}$ of pinned 
particles. In Ref.~\cite{CKD15}, $T_K$ was estimated by fitting the 
relaxation time $\tau_{\alpha}$ as a function of temperature  to the VFT 
formula for different values of $\rho_{pin}$.For many glass forming 
liquids without random pinning, $T_K$ turns out to be close to the 
extrapolated VFT divergence temperature $T_{VFT}$. It was observed 
~\cite{CKD15} that the $T_{VFT}$ obtained from the fits does not 
increase with increasing pin concentration. Instead, the kinetic fragility 
decreases drastically as $\rho_{pin}$ is increased. This behavior is in 
contradiction with the prediction in \cite{cammarotaPinning}. On the other 
hand, in Ref.~\cite{okim2015}, $T_K$ was estimated by calculating the 
configurational entropy (using a harmonic approximation for the basin
entropy) and it was found that $T_K$ increases with increasing pin 
concentration. Based on this observation, the existence of an ideal glass 
state beyond a ``critical'' pin concentration was predicted. However, it has 
been shown explicitly in Refs.~\cite{CKD15} and \cite{15CKDpnas} that the 
self overlap function $Q_s(t)$ and the self-intermediate scattering function 
decay to zero in time scales accessible in simulations even at state points 
where the configurational entropy is zero according to Ref~\cite{okim2015}. 
This proves that the $\alpha$ relaxation time remains finite~\cite{CKD15} 
at the ideal glass state points obtained in Ref.~\cite{okim2015}. This 
observation is in agreement with the finding in Ref.~\cite{CKD15} that 
$T_{VFT}$ does not increase with increasing $\rho_{pin}$. 

\begin{figure}
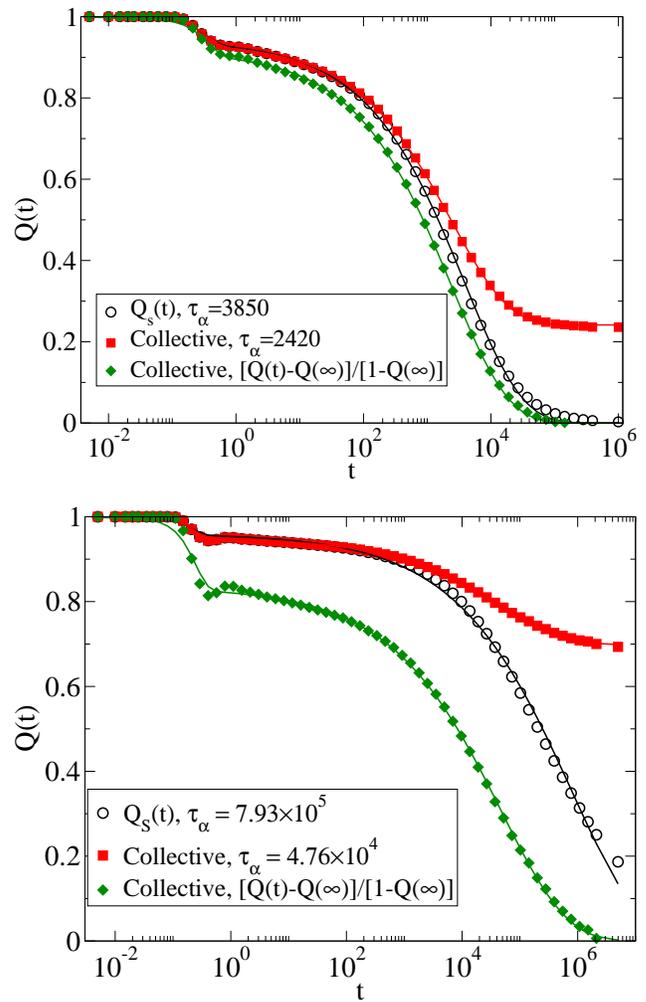

  \begin{center}
    \includegraphics[width=0.96\columnwidth]{{qtSelfCollective_0.600_1000_0.120}.eps}
    \vskip +0.3cm
    \includegraphics[width=0.96\columnwidth]{{qtSelfCollective_0.700_300_0.300}.eps}
    \caption{Full overlap correlation function and its self part for the 3dKA model 
at two state points close to the phase boundary of Ref.~\cite{okim2015}.
      {\em Top panel:} 3dKA, $N=1000,~T=0.600,~\rho_{pin}=0.120$.
      {\em Bottom panel:} 3dKA, $N=300,~T=0.700,~\rho_{pin}=0.300$. Solid lines
represent fits to the data points (see text for details).
      The time scale of decay of the full correlation function is shorter than
that for the self part.
    }
    \label{fullvsSelf}
  \end{center}
\end{figure}

A possible explanation of this discrepancy between the results for the 
dynamics and the configurational entropy was suggested in 
Ref.~\cite{okim2015PNASReply} where it was argued that one should extract 
the relaxation time from the
total density correlation function instead of its self part.
According to this argument, the total 
and self correlation functions decouple from each other as one 
increases the concentration of pinned particles and the relaxation time extracted from the 
total correlation function may diverge as $s_c$ goes to zero, even if
the time scale obtained from the self part remains finite. 
To test the validity of this argument, we have
calculated the relaxation time from the total overlap correlation function 
for the 3dKA model at state points close the phase boundary
in the $(\rho_{pin} - T)$ plane obtained in Ref.~\cite{okim2015}. 
In Fig.\ref{fullvsSelf}, we have 
%CD Need to specify the definition of the full overlap correlation function
%SK defined it along with the other definition.
plotted the full overlap correlation function and its self part for two state points 
near the phase boundary of Ref.~\cite{okim2015} and extracted relaxation times
from fits of the long-time data to a stretched exponential 
function. Since the full correlation function does not go to zero at long 
times (due to the presence of the random potential produced by the pinned particles), its long time value, $Q(\infty)$,
was obtained from the fits and the quantity $[Q(t) - Q(\infty)/[1 - Q(\infty)]$,
which decays from one to zero as $t$ is increased from zero to $\infty$, was used 
to obtain the relaxation time. The time scale obtained 
from the total overlap correlation function is found to be of the same order of
magnitude, {\em but smaller than} that obtained from the self part. A similar
conclusion was obtained from an analysis of the data for full and self correlation 
functions given in Ref.~\cite{okim2015PNASReply}. Thus, the discrepancy between the
dynamics and the results for $s_c(T)$ reported in \cite{okim2015} can not be
resolved by considering the time scale obtained from the full density correlation
function.

\begin{figure*}
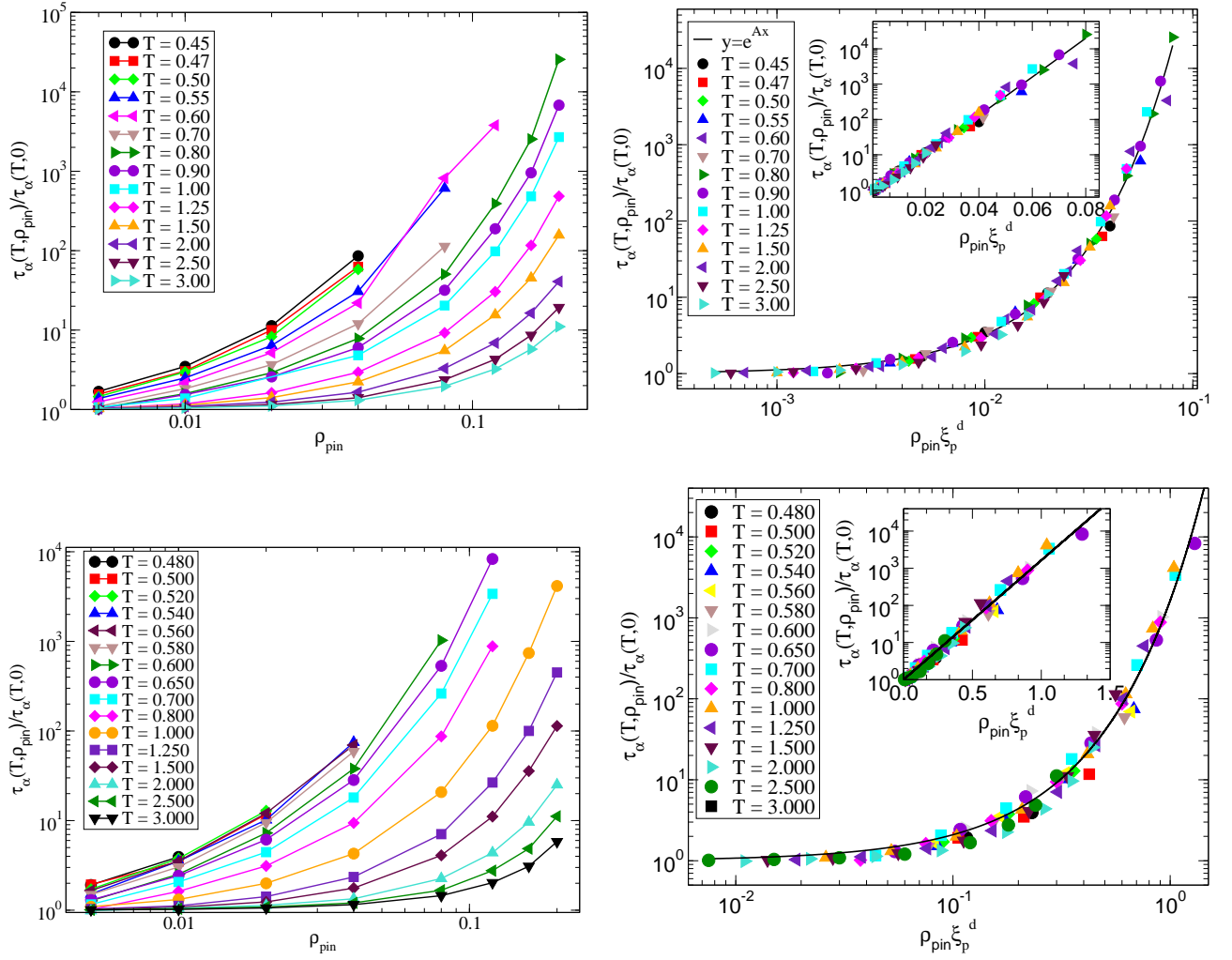

  \begin{center}
    \includegraphics[width=0.95\columnwidth]{noCollapse2dMKA.eps}\quad
    \includegraphics[width=0.98\columnwidth]{collapse2dMKAmod.eps}
    \vskip +0.4cm
    {\includegraphics[width=0.94\columnwidth]{updated_uncollapsed_Tau_2dR10.eps}
    \hskip +0.1cm\quad
    \includegraphics[width=0.97\columnwidth]{Updated_collapse_2dR10mod.eps}}
    \caption{Data collapse using the form in Eq. \ref{scalingfn} to obtain
      the length-scale $\xi_p$. The upper panels are for the 2dMKA system 
      and the lower panels are for the 2dR10 system. The left panels show
      the uncollapsed data. The lines passing through the collapsed data
      in both upper and lower right panels represent the scaling function 
      in Eq.\ref{scalingfn}. The scaling function provides a good description of 
      the collapsed data in both cases. Insets in both upper and lower panels
      represent the collapsed data in log-linear scale. The nice straight line 
      passing through the data clearly confirms our scaling arguments.}
    \label{collapse2dMKA}
  \end{center}
\end{figure*}

Here we will follow the arguments provided in Ref.~\cite{CKD15} for understanding
the dynamics of randomly pinned systems and use these arguments to analyze 
the simulation results obtained in the present study for
both two and three dimensional glass forming liquids. In our analysis,
we assume that: (a) the generalized AG relation of Eq.(\ref{modAG}) remains valid for systems with pinning; and (b)  $Ts^\alpha_c(T)$ goes to zero at the
Kauzmann temperature $T_K$ as $Ts^\alpha_c(T) \propto (T - T_K)$, so that the temperature dependence of the $\alpha$-relaxation time follows the VFT form. The dependence of $s_c$ on $T$ 
and $\rho_{pin}$ can be written as 
\begin{eqnarray}
Ts_c^{\alpha}(T,\rho_{pin}) &=& Ts_c^{\alpha}(T,0)F(T,\rho_{pin})\nonumber\\
                   &=&K(T-T_K) F(T,\rho_{pin}),
\label{neweq}
\end{eqnarray} 
where $F(T,\rho_{pin})$, the fractional reduction of the configurational 
entropy due to pinning, decreases from $1$ as $\rho_{pin}$ is increased 
from $0$. This is consistent with the general expectation of $s_c$ being 
a decreasing function of $\rho_{pin}$. Our observation that 
the VFT temperature is independent of
the value of $\rho_{pin}$ implies, via the generalized Adam-Gibbs relation, that 
$s_c$ goes to zero at the same temperature $T_K$ for
different values of $\rho_{pin}$. This, in turn, implies that for the small 
values of $\rho_{pin}$ considered here, $F(T,\rho_{pin})$ does not go to zero 
for temperatures higher than $T_K$. In Eq.~\ref{neweq}, $\alpha$ is the 
RFOT exponent defined in Eq.~(\ref{modAG})
and the second line follows from the expectation (verified in our simulations) that
the temperature dependence of $\tau_\alpha$ follows the VFT form. 
%If the temperature dependence of $F(T,\rho_{pin})$ is \textcolor{red}{assumed 
%to be} weak (we ignore the dependence of $F$ on $T$ in the following discussion), then 
Eq.( \ref{neweq}) and the generalized Adam-Gibbs relation lead to 
the following expression for $\tau_\alpha$ for non-zero $\rho_{pin}$:
\begin{equation}
  \tau_\alpha(T,\rho_{pin}) = \tau_\infty \exp\left[\frac{B(\rho_{pin})}{F(T,\rho_{pin})Ts_c^{\alpha}(T,0)}\right].
\label{tau-pinning}
\end{equation}
In Ref.\cite{15CKDpnas}, we have shown that 
$\ln[\tau_\alpha(T,\rho_{pin})]$
varies linearly with $1/[Ts_c^\alpha(T,0)]$ for all the values of $\rho_{pin}$ considered in our
study. This behavior implies that $F(T,\rho_{pin})$ is weakly dependent on or
independent of $T$. We ignore the dependence of $F$ on $T$ in the following discussion.
This equation predicts a VFT form 
for the temperature dependence of $\tau_\alpha$, with $T_{VFT} = T_K$ independent 
of $\rho_{pin}$, in agreement with the observations in Ref.~\cite{CKD15}. 
The fragility parameter $K_{VFT}= K T_{VFT} F(\rho_{pin})/B(\rho_{pin})$ 
would decrease with increasing $\rho_{pin}$ (i.e. would agree  
with our observations) if $F/B$ is a decreasing function of $\rho_{pin}$.   
This condition would be satisfied if $B$ increases, remains constant or 
decreases slower than $F$ with increasing $\rho_{pin}$. 

From Eq.(\ref{tau-pinning}), we get
\begin{equation} 
\ln \left[\frac{\tau_\alpha(T,\rho_{pin})}{\tau_\alpha(T,0)}\right] = G(\rho_{pin})/[Ts_c^\alpha(T,0)]
\label{gdef1}
\end{equation}
where 
\begin{equation}
G(\rho_{pin})\equiv \frac{B(\rho_{pin})}{F(\rho_{pin})} -  \frac{B(0)}{F(0)}.
\label{gdef2}
\end{equation}
Clearly, $G(0)$ is equal to zero and for small values of $\rho_{pin}$, we can write $G(\rho_{pin}) = C \rho_{pin} + C^\prime \rho_{pin}^2 + \cdots$, where $C$ and $C^\prime$ are
constants and $\cdots$ represent term with higher powers of $\rho_{pin}$. Using this in Eq.(\ref{gdef1}) above, we get
\begin{equation} 
\ln \left[\frac{\tau_\alpha(T,\rho_{pin})}{\tau_\alpha(T,0)}\right] = (C\rho_{pin} + C^\prime \rho_{pin}^2+\cdots)/[Ts_c^\alpha(T,0)].
\label{expansion}
\end{equation}
For small values of $\rho_{pin}$, we can retain only the term linear in $\rho_{pin}$ in the right-hand side of Eq.(\ref{expansion}) and obtain the following scaling relation: 
\begin{equation}
  \ln \left[\frac{\tau_\alpha(T,\rho_{pin})}{\tau_\alpha(T,0)}\right] = C f(\rho_{pin} \xi_p^d(T))
  \label{scalingfn}
\end{equation}
with  $f(x) = x$  and the pinning length scale $\xi_p$ is given by 
\begin{equation}
\xi_p(T)  = [1/(Ts_c^{\alpha}(T,0))]^{1/d} \propto [1/(T-T_{VFT})]^{1/d}
\label{xipdef}
\end{equation}
where $d$ is the spatial dimension. It is clear from its derivation that the scaling relation
of Eq.(\ref{scalingfn}) is valid only for small values of $\rho_{pin}$. As 
shown below, this scaling relation is satisfied by our numerical data in both 
two and three dimensions, indicating that for a fixed temperature $T$, the 
quantity $ \ln [\tau_\alpha(T,\rho_{pin})/\tau_\alpha(T,0)]$ is proportional 
to $\rho_{pin}$ for the small values of $\rho_{pin}$ considered in our work 
(see the insets of the right panels of Fig.~\ref{collapse2dMKA}).

The length scale $\xi_p$ can be identified as the average separation between neighboring pinned particles required for producing a 
given pinning-induced fractional change in the
relaxation time. Let us consider a small pinning-induced fractional change in the relaxation time, i.e. $\tau_\alpha(T,\rho_{pin})/\tau_\alpha(T,0)=1+x$ where the fractional change $x$ is a small positive number. 
Putting this in Eq.(\ref{scalingfn}), we get
\begin{equation}
\ln(1+x) \simeq x = C \rho_{pin} \xi_p^d(T),
\end{equation}
which can be written as 
\begin{equation}
\left(\frac{1}{\rho_p}\right)^{1/d}= \left( \frac{C}{x}\right)^{1/d}\xi_p(T)  .
\end{equation}
Since $(1/\rho_{pin})^{1/d}$ is proportional to the average distance between neighboring pinned particles, $\xi_p(T)$ is proportional to the average pin separation required
for producing a given fractional change $x$ in the relaxation time at temperature $T$. The increase of the length scale $\xi_p(T)$ with decreasing $T$ means that at lower temperatures,
a larger pin separation (smaller pin concentration) would be required for producing the same fractional change in the relaxation time. 
A very similar length scale with $\alpha =1$ was introduced 
in Ref.\cite{cammarotaPinning}, in which it was interpreted as 
``the critical mean distance between frozen particles''.
We are not aware of any correlation function whose spatial decay 
is governed by the length scale $\xi_p$.

This scaling prediction was tested and verified in Ref.~\cite{CKD15} for 
both the three dimensional models considered here. As shown in Fig. 
~\ref{collapse2dMKA}, it is possible to collapse all the data for 
$\psi(T,\rho_{pin}) \equiv 
\ln[\tau_\alpha(T,\rho_{pin})/\tau_\alpha(T,0)]$ for each of the two 
dimensional models into a single scaling curve by choosing the length scale 
$\xi_p(T)$ appropriately for different temperatures. The scaling function, 
represented by the line passing through the collapsed data points, is indeed 
of the form predicted by the scaling argument.

\begin{figure}
\begin{center}
\hskip -1.0cm
\includegraphics[width=1.0\columnwidth]{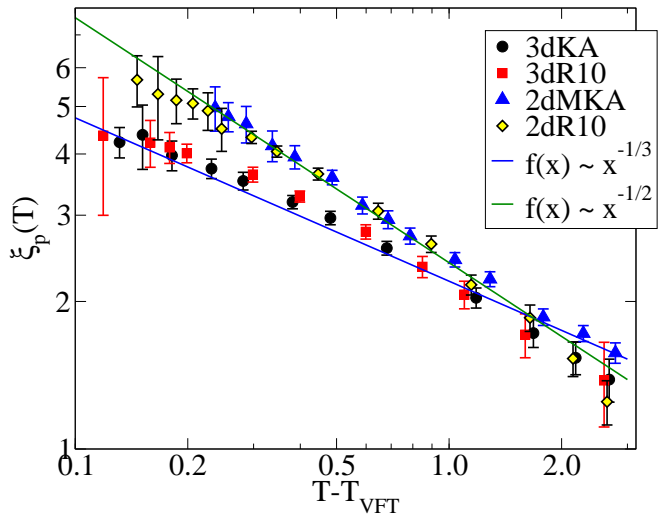}
\caption{$\xi_p$ vs $(T-T_{VFT})$ to demonstrate the validity of the 
scaling predictions.}
\label{xiPvsT}
\end{center}
\end{figure}
To check whether the $\xi_p(T)$ obtained from the scaling collapse is 
proportional to $ [1/(T-T_{VFT})]^{1/d}$, we have plotted in Fig. 
~\ref{xiPvsT} the pinning length scale $\xi_p$ obtained from the scaling 
collapse in Fig.~\ref{collapse2dMKA} as a function 
of $\left(T - T_{VFT}\right)$ for all the model systems. The results are 
clearly consistent with the predicted behavior. Thus, the scaling description 
proposed in Ref.~\cite{CKD15} provides a consistent description of the 
simulation data in both two and three dimensions. A very similar
scaling description has been used~\cite{expt2016} in a recent 
study to understand experimental results for the dynamics of a confined 
colloidal system.

\section{RFOT Exponents $\theta$ and $\psi$}

\begin{figure}
\begin{center}
%\vskip -0.67cm
\hskip -1.0cm
\includegraphics[width=1.0\columnwidth]{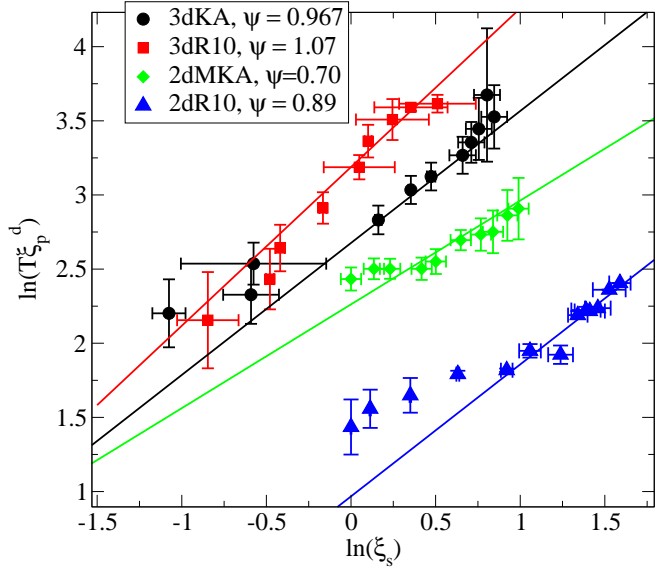}
\caption{$T\xi_p^d$ vs $\xi_s$ to extract the exponent $\psi$. Only the low 
temperature data points are fitted.} 
\label{xiSvsXip}
\end{center}
\end{figure}

\begin{figure}
\begin{center}
\vskip +0.5cm
\hskip -1.0cm
\includegraphics[width=1.05\columnwidth]{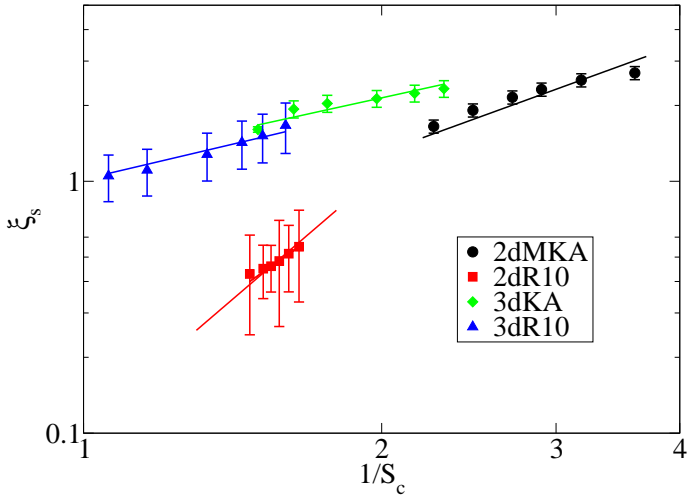}
\caption{The static length scale $\xi_s$ is plotted as a function of $1/s_c$  
for all the four model systems. The solid lines are fits to the form 
$\xi_s \propto \left(1/s_c\right)^{1/(d-\theta)}$ with $d-\theta$ obtained 
from the relation $\psi = \alpha (d - \theta)$. The values of $\psi$ are 
obtained from the fits shown in Fig.~\ref{xiSvsXip}, $\alpha=1$ for the three 
dimensional models and $\alpha$ for the two dimensional models are obtained 
from the fits shown in Fig.~\ref{modAG2dR10and2dmKA2}.} 
\label{xisSc}
\end{center}
\end{figure}

\begin{figure}
\begin{center}
%\hskip -0.5cm
{\includegraphics[scale=0.33]{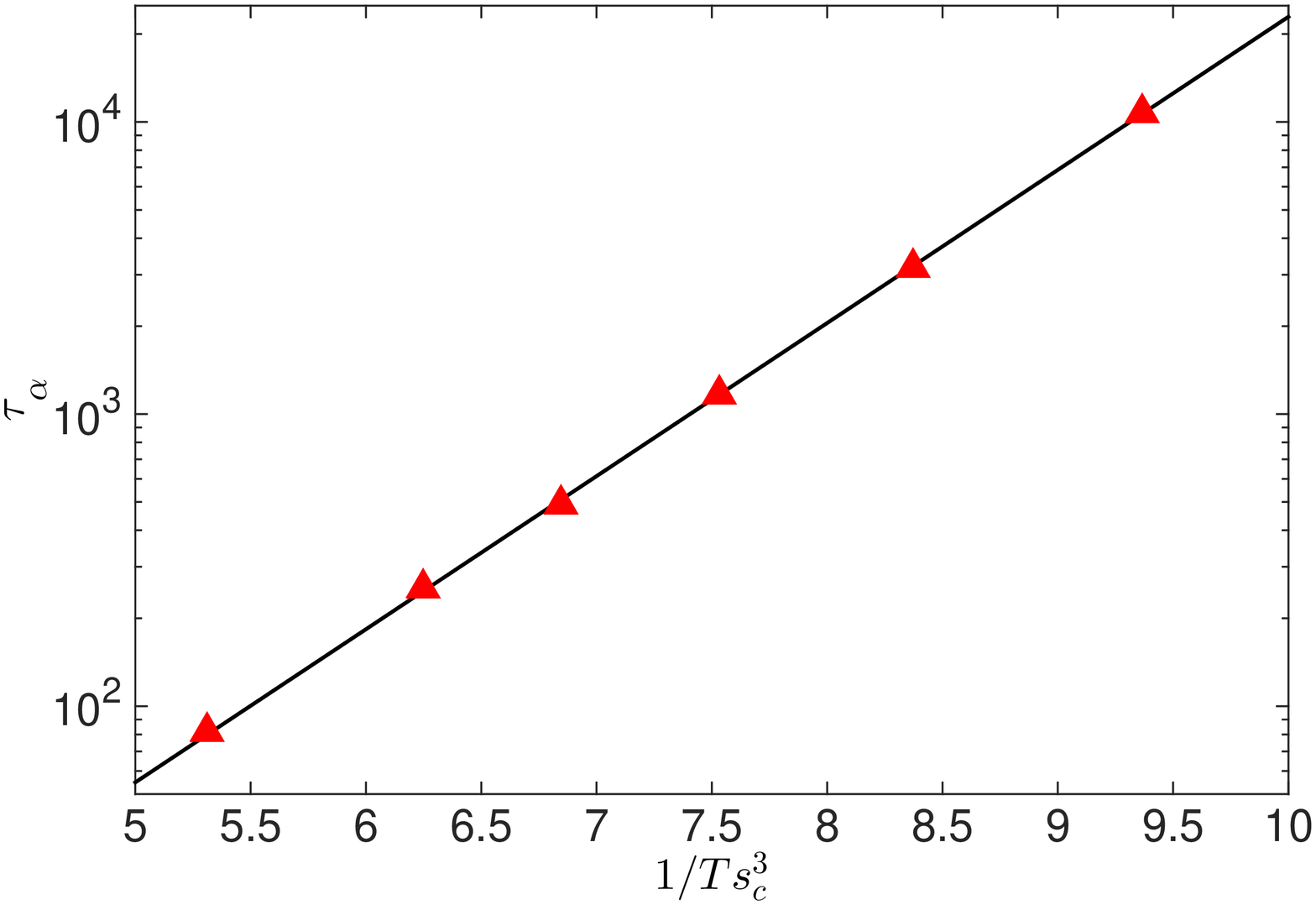}
\includegraphics[scale=0.35]{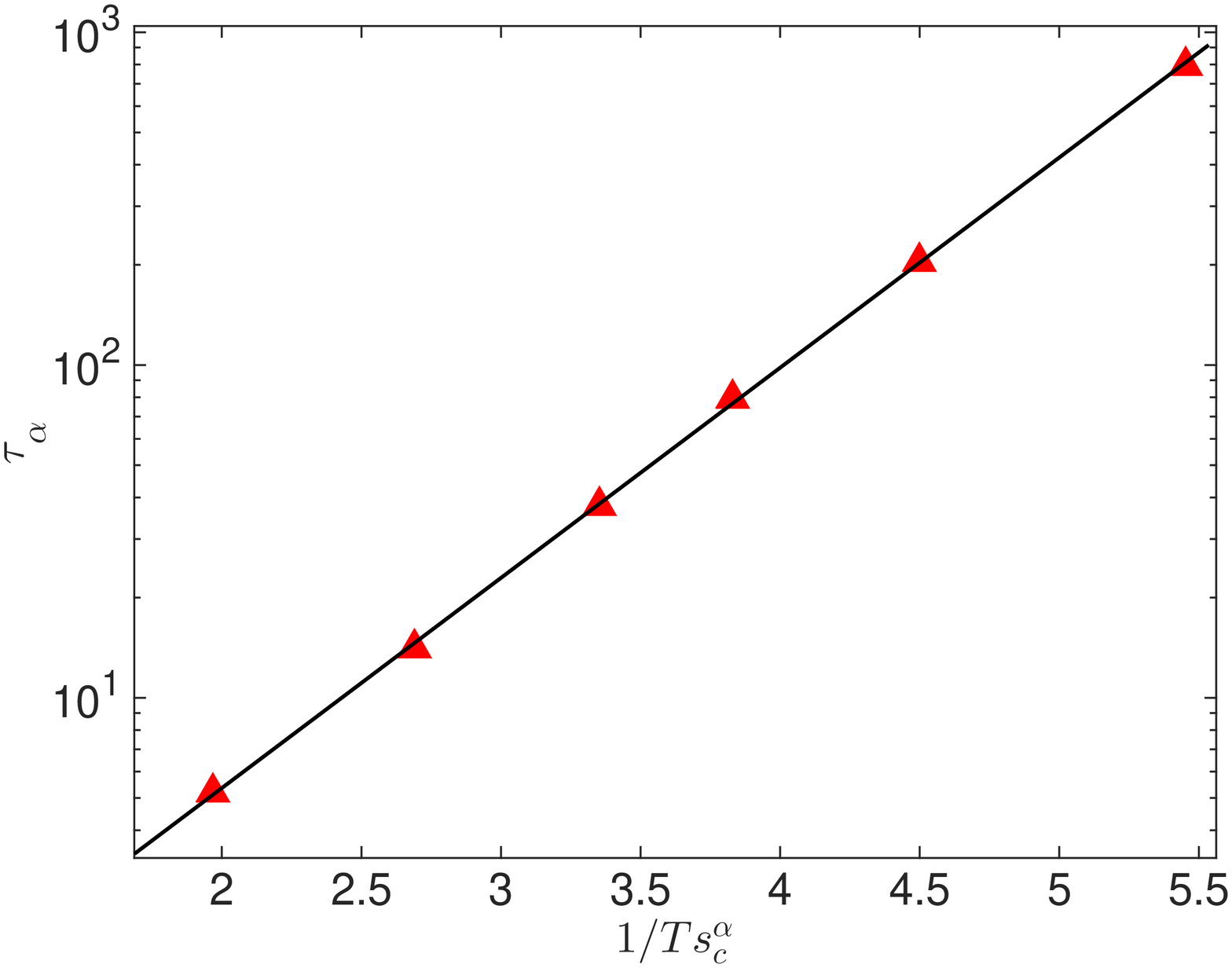}}
\caption{Top Panel: Modified Adam-Gibbs relation for the 2dR10 model.
Bottom Panel: Similar plot for the 2dMKA model with $\alpha \simeq = 0.70$.} 
\label{modAG2dR10and2dmKA2}
\end{center}
\end{figure}

\begin{figure}
\begin{center}
\hskip -0.5cm
\includegraphics[scale=0.40]{{T_TSc3_2dR10}.eps}
\includegraphics[scale=0.35, angle=-90]{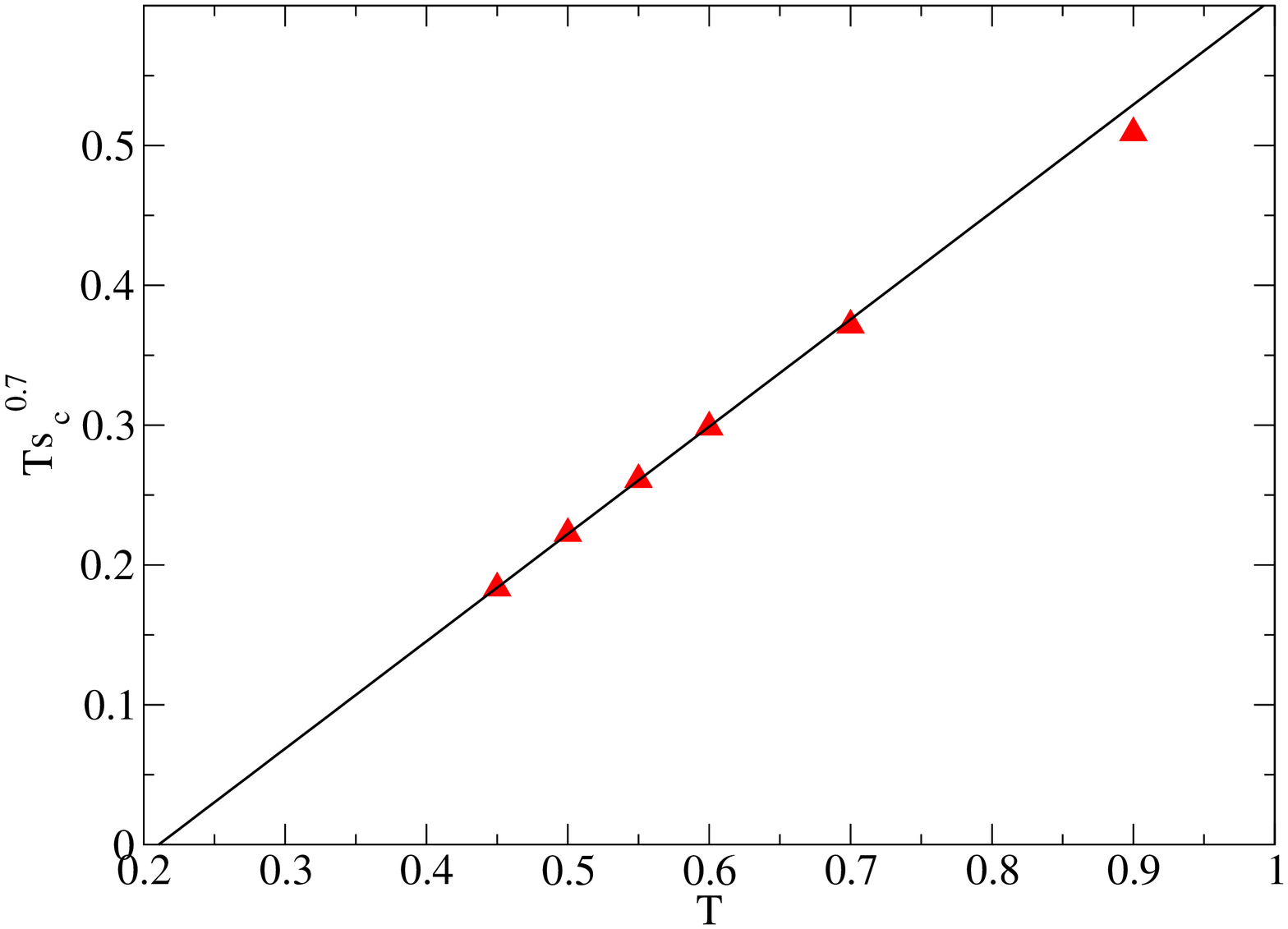}
\caption{Top Panel: $Ts_c^{\alpha}$ vs $T$ plot for the 2dR10 model
with $\alpha = 3.0$. Bottom Panel: Similar plot for the 2dMKA model with 
$\alpha = 0.70$. Fairly good straight line fits at lower temperatures
suggest that the relation $Ts_c^{\alpha} \propto T - T_K$ is satisfied, indicating that 
the VFT law will be obeyed.} 
\label{modAG2dR10and2dmKA1}
\end{center}
\end{figure}

In this section, we use the results for $\xi_p(T)$, $\xi_s(T,0)$, $s_c(T,0)$ and $\tau_\alpha(T,0)$
to estimate the values of the RFOT exponents $\theta$ and $\psi$ for the unpinned system. The exponent $\psi$ is obtained 
from the data for $\xi_p$ and $\xi_s$ (see Fig.\ref{xiSvsXip}). The exponent $\alpha=\psi/(d-\theta)$
for the two-dimensional models ($\alpha=1$ for the three-dimensional models) is obtained from the generalized AG relation (see Fig.~\ref{modAG2dR10and2dmKA2} - Fig.~\ref{modAG2dR10and2dmKA1} shows
that the values of $\alpha$ obtaind from the fits in Fig.~\ref{modAG2dR10and2dmKA1} are consistent
with the VFT form for the temperature dependence of $\tau_\alpha (T,0)$). The value of $\theta$ is
obtained from the calculated values of $\psi$ and $\alpha$. Finally, Fig.~\ref{xisSc}
provides a consistent check by examinimg whether the 
RFOT relation of Eq.(\ref{xis}) is satisfied for the calculated values of the exponent $\theta$.

The range of our data for the correlation lengths $\xi_s$ and $\xi_p$ is less than one decade. This is not sufficient
for an accurate determination of the power-law exponents. This difficulty in obtaining data 
over a range that is large enough to determine the RFOT exponents with high 
accuracy is intrinsic to the glass transition problem. Since the liquid falls 
out of equilibrium at temperatures substantially higher than the putative RFOT 
transition at $T_K$, the relevant correlation lengths $\xi_s$ and $\xi_p$ remain relatively small 
in the temperature range accessible in experiments or
simulations. For this reason, all existing attempts to determine the RFOT 
exponents from simulations~\cite{exponents1,exponents2} and experiments~\cite{exponents3}are based on data that span ranges 
comparable to or smaller than the ranges considered in our work. The novelty of our work is that we have used results for the
dynamics in the presence of pinning to estimate the values of these exponents.

%CD: check whether the caption of Fig.6 is correct
%SK: caption seems correct.

It follows from Eq.~\ref{xis} and Eq.~\ref{xipdef} that
\begin{equation}
T\xi_p^{d}(T) = \frac{1}{s_c^\alpha(T,0)} \propto \xi_s^\psi(T,0).
\label{xiSxiP}
\end{equation}
Thus, the data for $\xi_p(T)$ and $\xi_s(T,0)$ can be used to estimate the 
RFOT exponent $\psi$.  
In Fig.\ref{xiSvsXip}, $\ln(T\xi_p^d)$ is plotted {\it versus} $\ln(\xi_s)$ 
for all the four models considered here. The exponent $\psi$ obtained from 
the slope of this plot turns out to be close to unity for the two models 
studied in three dimensions. The best fit values, indicated by straight 
lines in the figure, are $\psi=0.967$ for the 3dKA model and $\psi=1.07$ for 
the 3dR10 model. The observation~\cite{ag-prl} that the AG relation is obeyed 
in both three dimensional models implies that $\alpha = \psi/(d-\theta)=1$ in 
three dimensions. This result, in combination of the result for $\psi$, imply
that $\theta \simeq 2.0$ in three dimensions. Our data for the dependence of 
$\xi_s$ on $s_c$ are also consistent with Eq.\ref{xis}. In Fig.\ref{xisSc}, 
$\xi_s$ is plotted as a function of $1/s_c$ for both the model systems and 
Eq.\ref{xis} with $(d-\theta)$ obtained from the relation $\psi = d - \theta$ 
provides a good description of the data for the three dimensional model 
systems. 

For the two dimensional models, the AG relation is violated~\cite{ag-prl} and
the relation $\alpha = \psi/(d-\theta)=1$ is not satisfied. We have obtained 
the exponent $\psi$ from Eq.\ref{xiSxiP} and $\psi/(d-\theta)$ from the
modified AG relation for the unpinned system. For the 2dMKA model, 
$\psi \simeq 0.70$ (see 
Fig.~\ref{xiSvsXip}), when only the low temperature data are considered and 
$\psi/(d-\theta) \simeq 0.70$ from the generalized AG plot shown in 
Fig.~\ref{modAG2dR10and2dmKA2}. Thus, $\theta \simeq 1.0$ for this model. In 
the $\xi_s$ vs $1/s_c$ plot in Fig.~\ref{xisSc}, $\theta \simeq 1.0$ describes 
the data quite well. Also, the relation $s_c^\alpha(T) \propto (T-T_K)$, 
required for the validity of the VFT form, is satisfied for $\alpha = 0.70$, 
as shown in Fig.~\ref{modAG2dR10and2dmKA1}.

For the 2dR10 model, $\psi \simeq 0.89$ from Fig.~\ref{xiSvsXip} if 
only the low temperature data are considered. A smaller value of $\psi$ 
($\sim 0.75$) is obtained if the data for a wider range of temperature are 
used in the power-law fit. The modified AG relation gives $\alpha = 
\psi/(d-\theta) \simeq 3.0$ (see Fig.~\ref{modAG2dR10and2dmKA2}).This 
implies $\theta \simeq 1.70$. The data for $\xi_s$ and $1/s_c$ are 
consistent with the expected relation between these two quantities
with $d - \theta \simeq 0.29$ (see Fig.\ref{xisSc}). Also, as shown in 
Fig.~\ref{modAG2dR10and2dmKA1}, the relation $s_c^\alpha(T) \propto (T-T_K)$ 
is satisfied by the data with $\alpha = 3.0$. However, the value of $\theta$ 
violates the physically expected condition, $\theta \leq d-1$.  Thus, although 
the exponent values $\psi \simeq 0.89$, $\alpha \simeq 3.0$, $\theta \simeq 
1.70$ provide an internally consistent description of the simulation data, 
the value of $\theta$ turns out to be unphysical. This unphysical value of 
$\theta$ is a consequence of the rather large value of $\alpha$.
For $\theta$ to be less than $(d-1)$ in two dimensions,
the exponents $\psi$ and $\alpha$ must satisfy the inequality $\alpha < \psi$. 
This is clearly ruled out by the data. The value of $\alpha$ reported in 
Ref.~\cite{ag-prl} for the 2dR10 model ($\alpha \simeq 2.1$) would also lead 
to a value of $\theta$ ($\simeq 1.5$) that is higher than $(d-1)$. 

Our results for the three-dimensional models ($\psi \simeq 1$ and $\theta \simeq 2$) are consistent with
those reported in Refs.~\cite{exponents1} and \cite{exponents2}. These values are different from those reported in 
Ref.\cite{09KDS} for the 3dKA model because the length scale considered in that work was the dynamic length scale which
is distinct from the static length scale $\xi_s$ considered here. The results for the two-dimensional models indicate that
the exponent values are not universal. It is also possible that the RFOT description breaks down in two dimensions.

%\textcolor{red}{Note that the exponents obtained in Ref.\cite{09KDS}
%where using the dynamic length scale as it was not clear then whether
%there exists both dynamic and static length scales in the problem.}

%\textcolor{red}{Previously the exponent $\psi$ were estimated from fitting 
%$log(\tau_{\alpha})$ vs $\xi_s/T$ with some exponents. For some of
%the works done on 2d systems by various groups (as reported in the 
%review Ref.[29]), a similar methodology was used and the range over which
%the relaxation time data were obtained was smaller than this work. 
%Also here we estimated the exponent $\psi$ by plotting $T\xi_p^d$ as
%a function of $\xi_s$ and then finding out the slope. This we believe
%is more controlled method for estimating the exponents than 
%the other previously employed methods. One of main highlights of this 
%present work is that one can use random pinning to get independent estimate
%of the exponent $\psi$. Later we used this estimated value of $\psi$
%to check the consistency of different relationships between these
%different exponents and our affirmative answer to this, suggest
%that the current estimate of the value of $\psi$ is probably more
%trustworthy than previous estimates. Although one should keep in 
%mind that the variation of both $\xi_p$ and $\xi_s$ are still not 
%large enough for very accurate estimation of different RFOT exponents 
%with the current computational power.}
 
\begin{figure*}
\begin{center}
\vspace{-0.4cm}
\includegraphics[width=0.99\columnwidth]{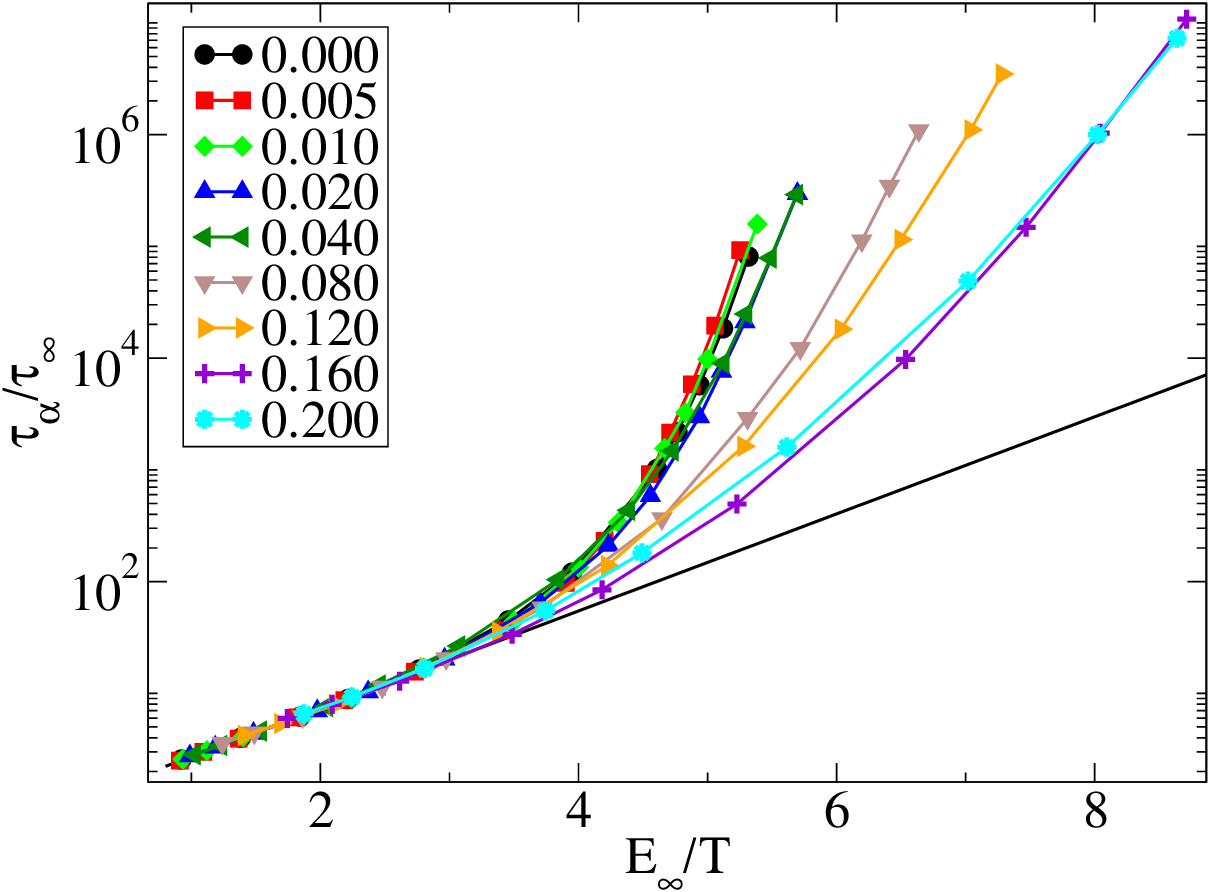}\qquad
\includegraphics[width=0.99\columnwidth]{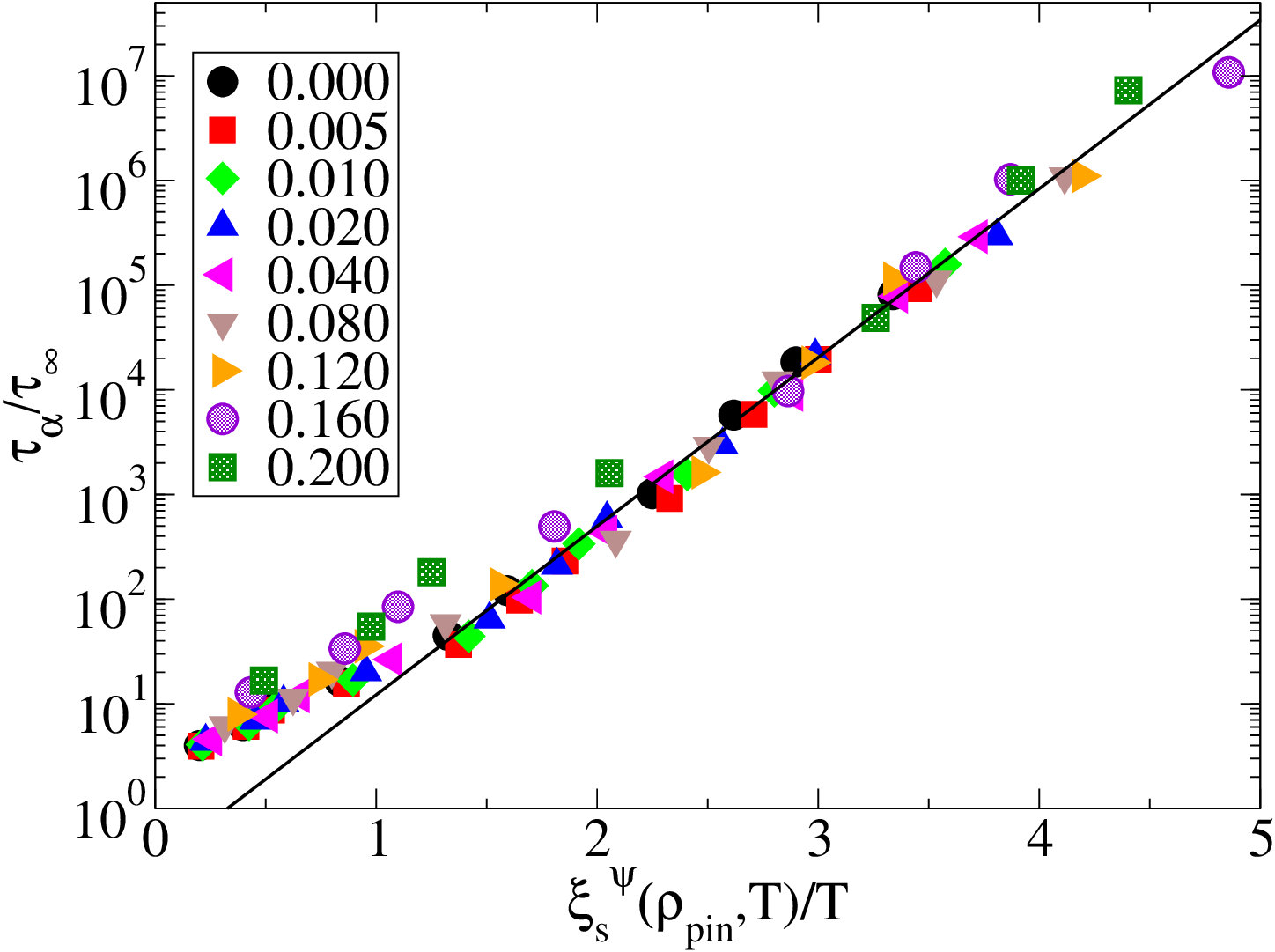}\\
\vskip +0.4cm
\includegraphics[width=0.99\columnwidth]{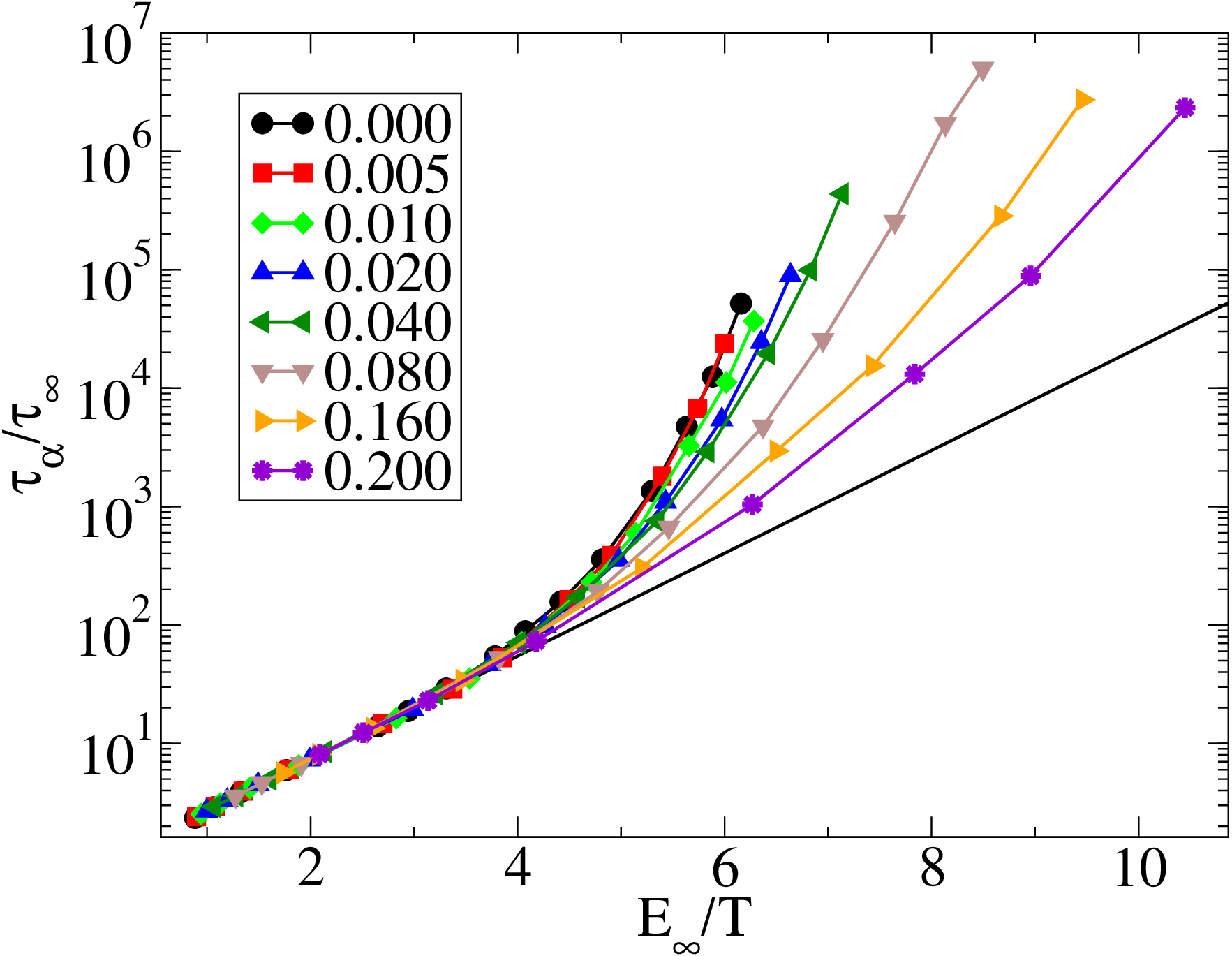}\qquad
%\qquad
\includegraphics[width=0.99\columnwidth]{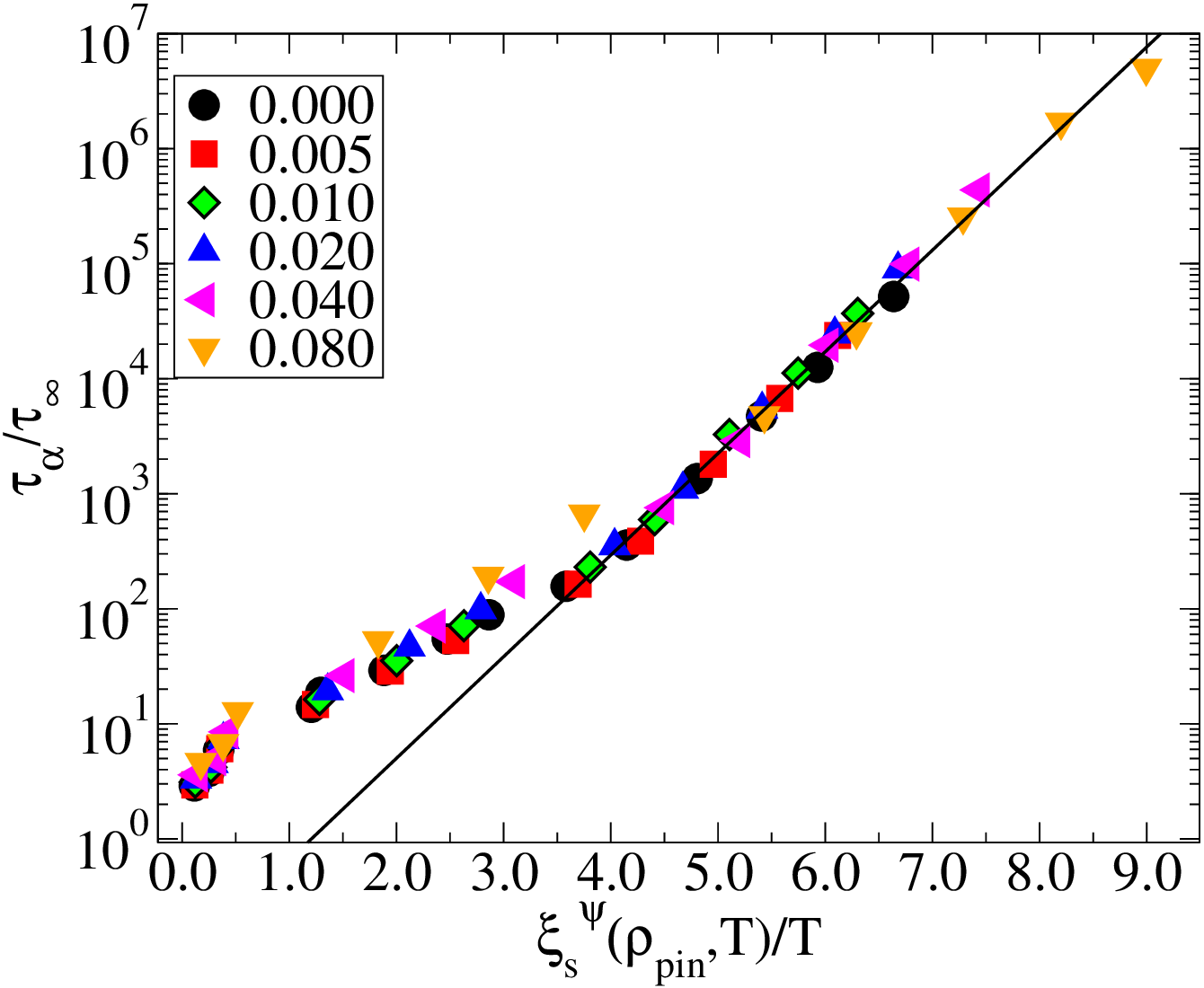}
\caption{{\em Top Left panel}: $\tau_\alpha/\tau_{\infty}$ versus 
$E_{\infty}/T$ for the 3dKA model with varying levels of pinning. Top Right 
panel: Data collapse using the static length scale $\xi_s(T,\rho_{pin})$. 
The data collapse is not good for higher levels of pinning since the scaling 
arguments presented in the text are valid in the weak pinning regime. Bottom 
panels: Similar analysis for the 3dR10 model. The values of $E_\infty$ and 
$\tau_\infty$ are provided in Table \ref{highTparameters}. For the 3dKA model,
$\psi=0.967$ and for the 3dR10 model, $\psi=$1.07.}
\label{fragilityCollapse3D}
\end{center}
\end{figure*}
%CD: These values of \psi are slightly different from those mentioned in the text. What is the reason for this?
%SK: these numbers were not updated during iteration of the paper. I fixed 
%them with the correct values. 
  
\section{Fragility and the Static Length Scale}
\begin{figure*}
\begin{center}
\includegraphics[width=0.98\columnwidth]{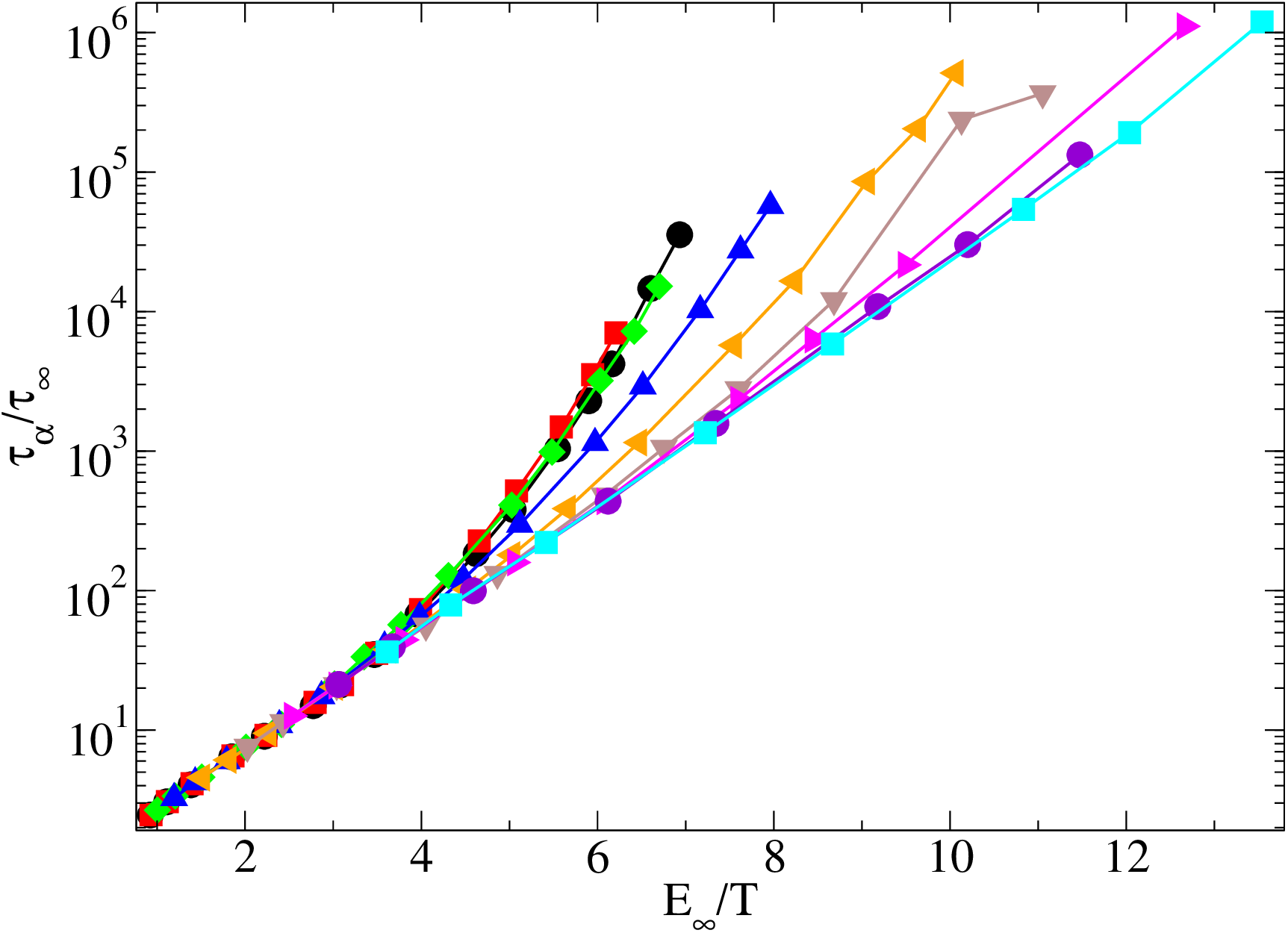}\qquad
\hskip +0.1cm
\includegraphics[width=0.95\columnwidth]{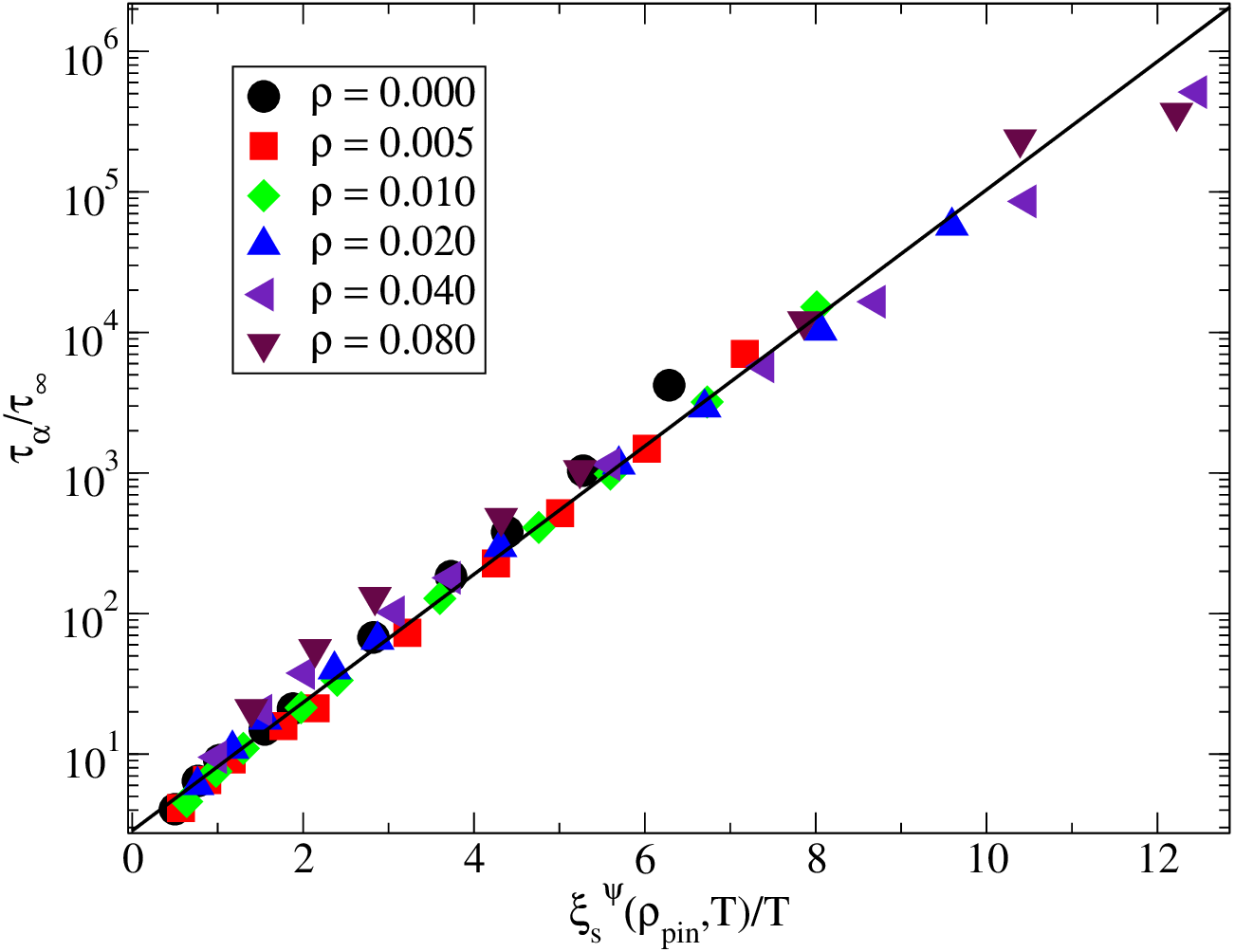}\\
\vskip +0.2cm
\includegraphics[width=0.97\columnwidth]{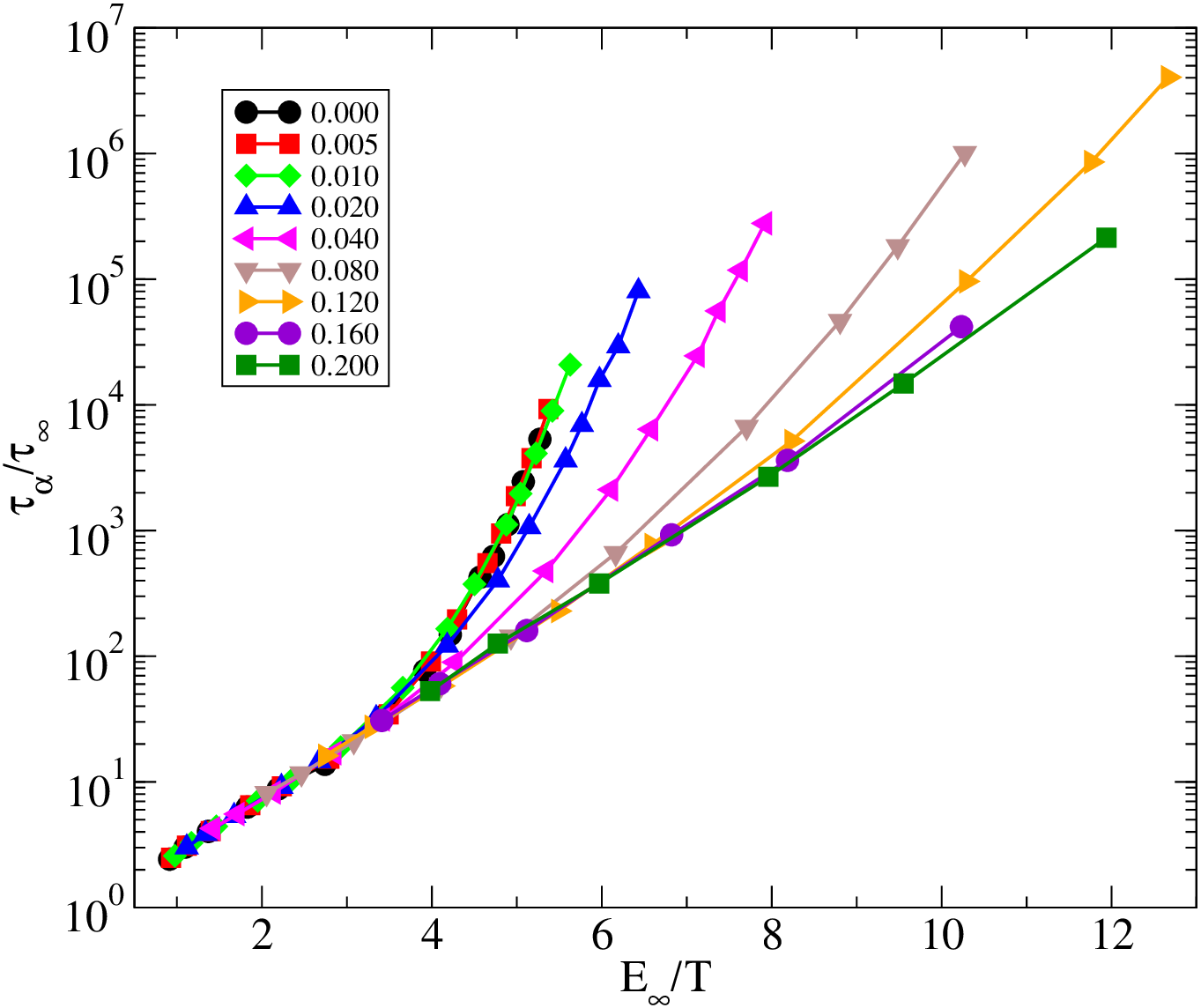}
%\qquad
\hskip 0.50cm
\includegraphics[width=1.02\columnwidth]{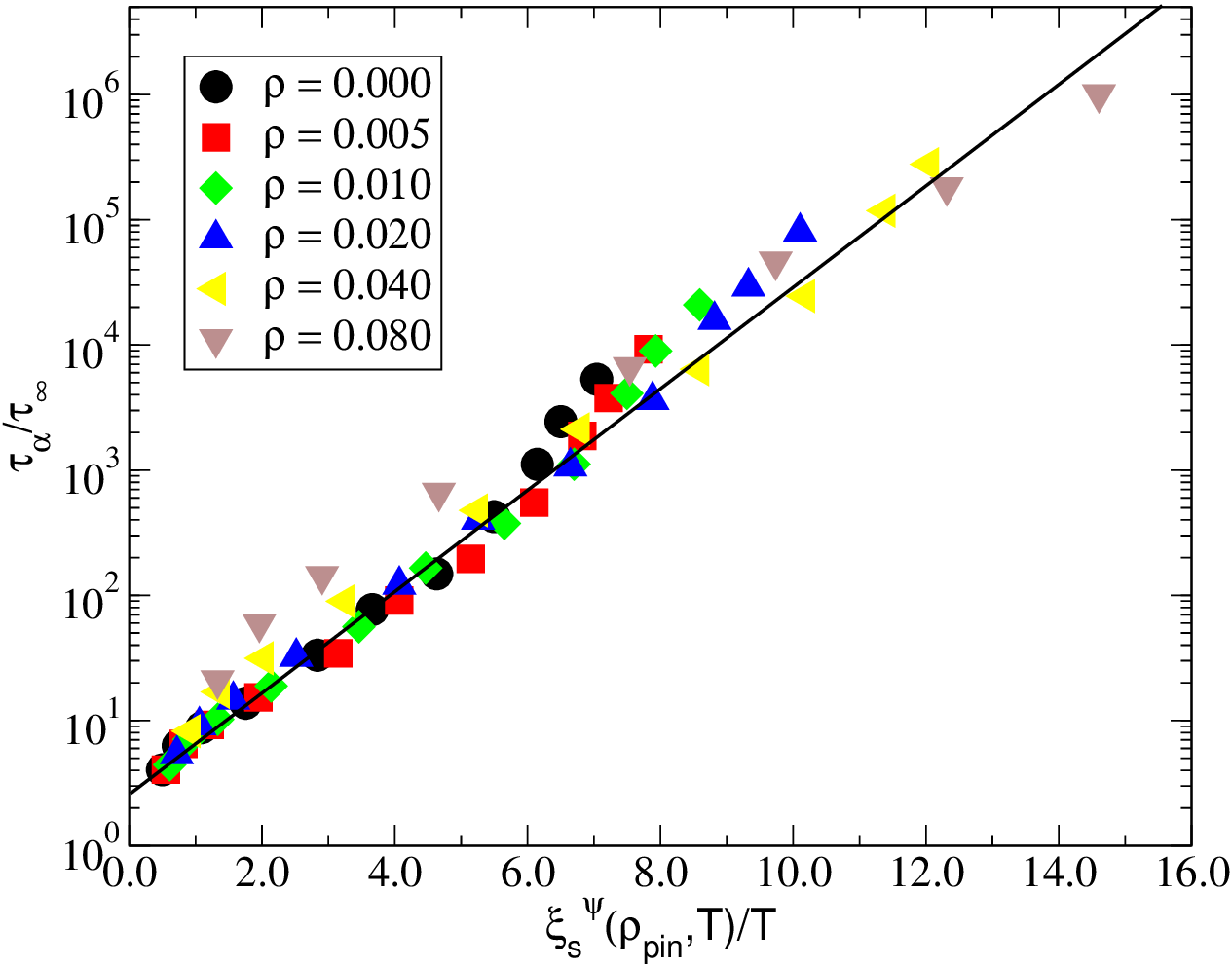}
\caption{{\em Top Left panel}: $\tau_\alpha/\tau_{\infty}$ versus 
$E_{\infty}/T$ for the 2dMKA model with varying levels of pinning.  
{\em Top Right panel}: Data collapse using the static length scale 
$\xi_s(T,\rho_{pin})$. The data collapse is not good for higher levels of
pinning since the scaling arguments presented in the text are valid in the
weak pinning regime. {\em Bottom panels:} Similar analysis for the 2dR10
model. The values of $E_\infty$ and $\tau_\infty$ are provided in Table 
\ref{highTparameters}. For the 2dMKA model, $\psi = 0.70$ and for the 2dR10 model, $\psi=0.89$.}
\label{fragilityCollapse2D}
\end{center}
\end{figure*}

As discussed in the Introduction, kinetic fragility measures the rate at which the relaxation time
changes with decreasing temperature. The role of fragility can be clearly demonstrated by 
plotting the relaxation time scaled by its infinite temperature value as 
a function of the temperature scaled by $E_{\infty}$, the
high temperature activation energy scale obtained by fitting the high 
temperature relaxation time data to the Arrhenius form,
\begin{equation}
\frac{\tau_{\alpha}(T)}{\tau_{\infty}} = \exp\left({\frac{E_{\infty}}{T}}\right).
\label{arrh}
\end{equation}
This way of plotting the relaxation time data for different systems makes
sure that the high temperature part of the data for all the 
systems fall on a single master curve which would be a
straight line if  $\ln({\tau_{\alpha}/\tau_{\infty}})$ is plotted as a function
of $E_{\infty}/T$. Deviations from this straight line at lower
temperatures would highlight the varying degree of fragility of these 
different systems. In the left panels of Fig.~\ref{fragilityCollapse3D} and 
Fig.~\ref{fragilityCollapse2D}, 
one can clearly see that the relaxation time data at high temperatures
fall on a straight line. Values of the parameters $E_{\infty}$ and $\tau_{\infty}$,
obtained from fits of the high-temperature data to the Arrhenius form of Eq.~\ref{arrh}
are given in Table~\ref{highTparameters} for all the models considered here. 
The data for systems with different 
values of $\rho_{pin}$ deviate from this straight line by different amounts at
low temperatures. The unpinned 
system deviates the most and the system with $\rho_{pin} = 0.20$ shows the 
least amount of deviation, illustrating a decrease of fragility with increasing 
pin density. 
\begin{table}[!h]
\begin{tabular}{| c | c | c | c | c | c | c | c | c |}
\hline
 &\multicolumn{2}{c|}{3dKA} &\multicolumn{2}{c|}{3dR10}
 &\multicolumn{2}{c|}{2dMKA} &\multicolumn{2}{c|}{2dR10}\\
\hline
$\rho_{pin}$ & $E_\infty$ & $\tau_\infty$ & $E_\infty$ & $\tau_\infty$
& $E_\infty$ & $\tau_\infty$ & $E_\infty$ & $\tau_\infty$\\
\hline
0.000 & 2.648 & 0.133 & 2.765 & 0.093 & 2.773 & 0.280 & 2.741 & 0.213
\\
0.005 & 2.698 & 0.130 & 2.731 & 0.095 & 2.794 & 0.284 & 2.793 & 0.209
\\
0.010 & 2.828 & 0.124 & 2.800 & 0.093 & 3.015 & 0.271 & 2.926 & 0.204
\\
0.020 & 2.986 & 0.120 & 2.962 & 0.088 & 3.583 & 0.234 & 3.345 & 0.181
\\
0.040 & 3.210 & 0.115 & 3.073 & 0.087 & 4.528 & 0.198 & 4.277 & 0.140
\\
0.080 & 3.823 & 0.103 & 3.717 & 0.072 & 6.078 & 0.178 & 6.164 & 0.092
\\
0.120 & -- & -- & 4.224 & 0.064 & 7.600 & 0.177 & 8.233 & 0.065 \\
0.160 & 5.201 & 0.084 & 5.228 & 0.048 & 9.180 & 0.187 & 10.233 & 0.052
\\
0.200 & 6.270 & 0.070 & 5.615 & 0.047 & 10.834 & 0.210 & 11.939 &
0.057 \\
\hline
\end{tabular}
\caption{Values of $E_\infty$ and $\tau_\infty$ for different pin concentrations,
obtained from the fits shown in Figs.~\ref{fragilityCollapse3D} and \ref{fragilityCollapse2D}.}
\label{highTparameters}
\end{table}

If the relaxation time $\tau_{\alpha}$ is  determined solely by the static 
length scale $\xi_s$, as in Eq.~\ref{tauXi} with $\Delta$ a system dependent 
constant, then one would expect that the fragility will also be completely 
determined by the growth of the static length scale with decreasing temperature. 
If the exponent $\psi$ is the same for systems with different pin densities, 
then plots of the scaled relaxation time as a function of 
$\xi_s^{\psi}(T,\rho_{pin})/T$ should collapse to a straight line at low 
temperatures. This possibility is checked below by approximately calculating 
the static length scale for different pin densities. The objective here is 
to examine whether the observed fragility change with increasing pin 
concentration can be understood as an effect of a pinning-induced change in 
the temperature dependence of the static length scale. 

We start from our proposal for the dependence of the configurational entropy on
pin concentration, Eq.\ref{neweq}. For small pin concentrations, 
$\rho_{pin}\to0$, we can express $F(\rho_{pin})$ as $F(\rho_{pin})= 1-
C\rho_{pin} + \ldots\,$, $C>0$ and the corresponding configurational entropy as 
\begin{equation}
s_c(T,\rho_{pin}) = s_c(T,0)(1-C\rho_{pin}).
\end{equation}
Now, assuming that the RFOT relation between the static length scale and the
configurational entropy (Eq.~\ref{xis}) remains valid for pinned systems, 
we can write
\begin{eqnarray}
\xi_s(T,\rho_{pin}) &\propto& \left[\frac{1}{Ts_c(T,\rho_{pin})}\right]^
{1/(d-\theta)}\nonumber\\
&=& D(\rho_{pin})\left[\frac{1}{Ts_c(T,0)}\right]^\frac{1}{d-\theta}
\end{eqnarray}
with $D(\rho_{pin}) = 1 + A\rho_{pin} + \ldots \,\,$, $A>0$. Thus, we have
\begin{eqnarray}
\xi_s(T,\rho_{pin})=\xi_s(T,0)(1+A\rho_{pin}).
\label{xiRhoRelationFromRfot}
\end{eqnarray}
As discussed earlier, the static length scale for $\rho_{pin} = 0$ for the 
three dimensional models is obtained from a combination of finite size scaling 
analysis of the minimum eigenvalue of the Hessian matrix and PTS analysis (see 
\cite{BKP13PRL} for further details). Finite size scaling of $\tau_{\alpha}$ 
is used to extract the static length scale for the two dimensional model 
systems studied here. Once the length scale at zero pinning is obtained,
Eq.\ref{xiRhoRelationFromRfot} is used to obtain the static length scale for 
different pin densities, treating $A$ as a system dependent constant. 
This length scale is then used to obtain the data collapse in 
Figs.~\ref{fragilityCollapse3D} and \ref{fragilityCollapse2D}. The value of 
$A$ for each model system is determined by maximizing the quality of data 
collapse. The collapse of the $\tau_\alpha$ data for varying levels of 
pinning is observed to be quite good within our numerical accuracy. The data 
for the larger values of $\rho_{pin}$, $\rho_{pin}\in\{16\%,20\%\}$ could not 
be collapsed for the 3dKA model, suggesting that in this case, one may need to 
include higher order terms in Eq.\ref{xiRhoRelationFromRfot}. This analysis 
suggests a deep connection between the static length scale and the fragility 
in different glass forming liquids.

\section{Conclusions}
In this article we have presented results from extensive computer simulations 
of the dynamics of several model glass forming liquids in the presence of 
random pinning in both two and three dimensions. The effects of random pinning 
on the dynamics of two dimensional systems are found to be qualitatively 
similar to those reported earlier \cite{CKD15} for three dimensional model 
glass formers. The presence of random pinning does not lead to an increase in 
the VFT temperature $T_{VFT}$, but increases the kinetic fragility of the 
liquids, irrespective of the model and the spatial dimension considered in the 
simulation. We have tried to understand the observed behavior within the 
framework of the RFOT theory and provided a scaling description that explains 
our simulation results for the dependence of the $\alpha$ relaxation time and 
the fragility on the pin concentration.

The main results of this work are:
\begin{itemize}
\item{The fragility of a glass forming liquid can be tuned systematically 
over a large range by pinning a randomly chosen subset of the particles in 
their equilibrium positions. This effect is generic for glass forming liquids 
in both two and three dimensions.}

\item{We find that pinning induced changes in the temperature dependence 
of the $\alpha$ relaxation time $\tau_\alpha$ can be explained entirely in 
terms of the dependence of the static length scale of RFOT, 
$\xi_s(T,\rho_{pin})$, on the pin concentration $\rho_{pin}$. 
This observation implies that the kinetic fragility
is intimately connected with the static length scale. In particular, the  
RFOT relationship between $\tau_{\alpha}$ and $\xi_s$, 
$\tau_{\alpha} = \tau_0 \exp{\left(\Delta \xi_s^{\psi}/T\right)}$
describes the data quite well and appears to be universal
in nature.}

\item{A scaling description, consistent with the RFOT theory and based on the
assumption that random pinning changes the configurational entropy by a 
multiplicative factor that decreases from unity as the pin concentration is 
increased from zero, provides a good description of all simulation results for 
the dynamics. This description reveals the existence of a pinning-related 
static length scale, $\xi_p$. The dynamics of supercooled liquids with random 
pinning can be completely understood in terms of this length scale and the 
other static length scale, $\xi_s$. A scaling argument suggests the existence 
of a simple relation between these two length scales and gives us an 
opportunity to extract one of the exponents (the barrier height exponent 
$\psi$) of RFOT theory from the simulation data.}

\item{Combining the results for $\psi$ with those for the exponent $\alpha$ 
in the generalized AG relation, we can extract the value of the RFOT surface 
tension exponent $\theta$. Our results for the three dimensional models are 
consistent with $\theta = (d-1)$, $\psi = 1$. In one of the two dimensional 
models (the 2dMKA model), we find $\theta \simeq d-1$, $\psi \simeq 0.7$. In 
the other two dimensional model (the 2dR10 model), the value of $\psi$ is 
found to be close to 1, but the value of $\theta > 1.0$ violates the 
inequality $\theta \leq (d-1)$. This suggests that the RFOT description is 
not applicable to this system. Our results for the two dimensional systems 
are consistent with earlier observation~\cite{ag-prl} of breakdown of the AG 
relation and lack of universality in two dimensions. This may be related to 
recent results~\cite{szamelNatComm15} suggesting that the behavior of glass 
forming liquids in two dimensions is qualitatively different from that 
observed in three dimensions. }

\item{ Relations among the static length scale $\xi_s$, the configurational 
entropy $s_c$ and the $\alpha$ relaxation time $\tau_\alpha$ predicted in the 
RFOT theory are satisfied by our simulation results with the obtained exponent 
values.}

\end{itemize}

In conclusion, this study clearly demonstrate that many of the 
simulation results for the dynamics of different glass forming liquids in 
the presence of random pinning can be understood within the framework of the 
RFOT theory and random pinning provides an interesting tool to unravel several 
aspects of glassy dynamics. More simulational and experimental research along 
these lines will be very useful.  

%CD need to put all citations in the same format.
%SK done

\bibliography{$HOME/Dropbox/docs/jShort.bib,$HOME/Dropbox/docs/mybiblio.bib}

\end{document}